\documentclass[openacc]{rstransa}
\usepackage{multirow}
\usepackage{colortbl}
\usepackage{xcolor}
\usepackage{soul}
\usepackage{verbatim} 
\usepackage{makecell}
\usepackage{todonotes}
\usepackage{anyfontsize}
\usepackage{float}
\usepackage{tabularray}
\usepackage{pdflscape}
\usepackage{booktabs}
\usepackage{rotating}
\usepackage{hyperref}
\usepackage{tikz}
\usepackage{orcidlink}

\makeatletter
\def\@doi{10.1098/rsta.2023.0393}

\renewcommand{\@history}{%
    {\fontfamily{\sfdefault}\fontseries{m}\fontshape{n}\fontsize{10}{14}\selectfont\raggedright
    Preprint. Article submitted to journal.\par
    DOI: \href{https://doi.org/\@doi}{\@doi}\par}}
\makeatother

\begin{document}

\title{IMPACT : \textbf{I}n-\textbf{M}emory Com\textbf{P}uting Architecture Based on Y-Fl\textbf{A}sh Technology for \textbf{C}oalesced \textbf{T}setlin Machine Inference}

\author{
Omar Ghazal\orcidlink{0000-0001-6742-9600}$^{1}$ ,Wei Wang$^{2}$, Shahar Kvatinsky$^{3}$, Farhad Merchant$^{1,4}$, Alex Yakovlev$^{1}$ and Rishad Shafik\orcidlink {0000-0001-5444-537X}$^{1}$}

\address{$^{1}$Microsystems Group, School of Engineering
Newcastle University, NE1 7RU, UK.\\
$^{2}$Peng Cheng Laboratory, Shenzhen, China.\\
$^{3}$Technion—Israel Institute of Technology, Haifa, Israel.\\
$^{4}$Bernoulli Institute and CogniGron, the Netherlands.}

\subject{artificial intelligence, pattern recognition, electrical engineering, microsystems}


\corres{Omar Ghazal\\
\email{o.g.g.awf2@newcastle.ac.uk}}

\begin{abstract} The increasing demand for processing large volumes of data for machine learning models has pushed data bandwidth requirements beyond the capability of traditional von Neumann architecture. In-memory computing (IMC) has recently emerged as a promising solution to address this gap by enabling distributed data storage and processing at the micro-architectural level, significantly reducing both latency and energy. In this paper, we present the IMPACT (\textbf{I}n-\textbf{M}emory com\textbf{P}uting architecture based on Y-Fl\textbf{A}sh technology for \textbf{C}oalesced \textbf{T}setlin machine inference), underpinned on a cutting-edge memory device, Y-Flash, fabricated on a 180 nm CMOS process. Y-Flash devices have recently been demonstrated for digital and analog memory applications, offering high yield, non-volatility, and low power consumption. The IMPACT leverages the Y-Flash array to implement the inference of a novel machine learning algorithm: coalesced Tsetlin machine (CoTM) based on propositional logic. CoTM utilizes Tsetlin automata (TA) to create Boolean feature selections stochastically across parallel clauses. The IMPACT is organized into two computational crossbars for storing the TA and weights. Through validation on the MNIST dataset, IMPACT achieved \(96.3\%\) accuracy. The IMPACT demonstrated improvements in energy efficiency, e.g., \(2.23X\) over CNN-based ReRAM, \(2.46X\) over Neuromorphic using NOR-Flash, and \(2.06X\) over DNN-based PCM, suited for modern ML inference applications. 
\end{abstract}



\begin{fmtext}

\end{fmtext}


\maketitle
\section{Introduction} In machine learning (ML) systems, algorithm architectures are inherently driven by data during both the training and inference regimes. As such, they are confronted with challenges related to data storage, communication, processing, and the associated energy costs. These issues are exacerbated further by the growing demands of ML applications with large dimensionality of data. The traditional von Neumann computing architectures, which separate processing and memory units, do not scale well for such data-driven workloads. These architectures incur larger latency and increased energy consumption when fetching the data between the CPU and memory, thereby limiting the computational efficiency~\cite{sebastian2020memory,krestinskaya2024neural}. 

In-memory computing (IMC) offers a promising solution to address these challenges by integrating storage and processing of data. Compared to von Neumann architectures, this approach significantly reduces the data transfer overheads as well as offers natural parallelization between different processing units. Motivated by these advantages, over the years, several IMC approaches leveraging digital and analog memory technologies were proposed~\cite{sebastian2020memory, christensen20222022, krestinskaya2024neural,10473924}.

Traditional digital-based complementary metal oxide semiconductor (CMOS) memories, including those based on static random access memory (SRAM)  and dynamic random access memory (DRAM), have been investigated for IMC architectures~\cite{krestinskaya2024neural,sebastian2020memory}. These memory arrays can be reconfigured to perform bit-wise (dot multiplication) operations, making them suitable for parallel data processing. Utilizing digital memory arrays for IMC offers advantages such as established manufacturing processes, compatibility with existing digital logic, and high-density integration on a single chip~\cite{ISSCC_10067643}. However, the digital memory arrays have major limitations, including substantial static leakage, frequent logic transitions, and refreshing that cause an increase in both energy consumption and latency. Moreover, implementing complex arithmetic operations within memory arrays requires specialized hardware design and complex peripheral circuitry for computational tasks~\cite{yu2024ferroelectric, sebastian2020memory, krestinskaya2024neural}.

To address the above limitations, emerging non-volatile analog memory devices, notably memristors, have been explored. Examples include resistive RAM (ReRAM)~\cite{wan2022computeRERAM}, phase-change memory (PCM)~\cite{sebastian2019computationalPCM}, and magnetoresistive RAM (MRAM)~\cite{jung2022crossbarMRAM}. Memristors can adjust its resistance based on the applied voltage/current; they can process and store information simultaneously, allowing for more efficient and parallel data processing across multiple storage units~\cite{Yu_9116417}. These devices can store information in a Boolean manner, similar to the two distinct binary states of digital memory, with the highest conductance state (HCS) or the lowest conductance state (LCS). Moreover, they can store information in an analog fashion, representing a continuous range of values, mimicking the behavior of synapses in the human brain~\cite{serb2016unsupervised}. These distinguishing characteristics make memristors well-suited for applications in neuromorphic computing and various machine learning algorithms, including novel approaches like the Tsetlin machine \cite{2018-granmo-tsetlin}, where energy consumption and processing speed are core design objectives~\cite{bao2022toward}.


Despite their potential, memristors have drawbacks that delay their widespread implementation in ML applications. One of the main challenges is their variability; memristors can exhibit inherent variations in their resistance states, leading to noise in computations and potentially impacting accuracy. Furthermore, memristors' endurance and retention properties are still active research areas, as these factors can affect the long-term reliability and stability of the devices~\cite{mannocci2023memory,lanza2021standards}.

Recently, a novel non-volatile memristor device known as Y-Flash was introduced in~\cite{YFLASH2022_DBNN,YFLASH2019_2terminal,YFLASHMODEL_physical}. These memristors are designed to be compatible with CMOS processes. The Y-Flash device features a floating gate isolated from its two sources, with the control gate merged with the drain. This unique design significantly reduces the device footprint while enabling program, erase, and readout operations in a two-terminal configuration. They offer the precision of digital memory and the continuous range of analog memory. In other words, the devices can operate in both digital modes (LCS and HCS) and can be fine-tuned within these states $(1\ nA - 5\ \mu A)$, providing analog-like functionality. When organized in a crossbar array, the Y-Flash devices can eliminate sneak-path currents, thus removing the need for the selector device. This hybrid memristor shows promise in addressing the limitations of purely digital or analog memory and offers a flexible and efficient platform for IMC architectures designed for ML applications ~\cite{YFLASH2022_DBNN,YFLASH2019_2terminal,YFLASHMODEL_physical}.

\begin{figure}[t]
\centering\includegraphics[width=\linewidth]{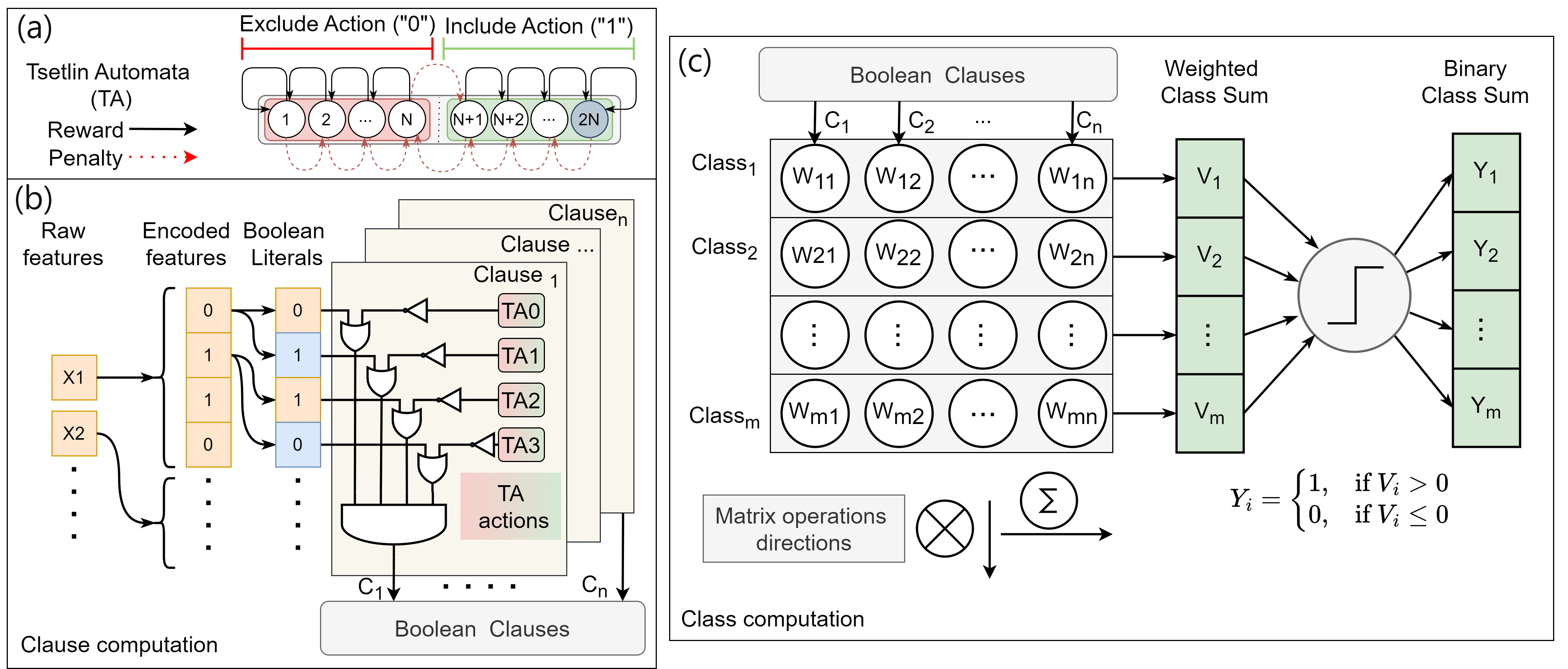}
\caption{The coalesced Tsetlin machine algorithm data pipeline consisting of (a) the learning element of CoTM, Tsetlin Automata (TA), (b) the interaction of the TA with the related Boolean literals forms the clauses, and  (c) the clauses' weight matrix to produce the class output vector (Y).}
\label{fig_CoTM_Algo}
\end{figure}

This paper presents an efficient IMC architecture, named IMPACT (\textbf{I}n-\textbf{M}emory com\textbf{P}uting architecture based on Y-Fl\textbf{A}sh technology for \textbf{C}oalesced \textbf{T}setlin machine inference). The  IMPACT employing a Y-Flash array to enhance the inference capabilities of a novel machine learning algorithm called the coalesced Tsetlin machine (CoTM) (see Figure~\ref{fig_CoTM_Algo})~\cite{glimsdal2021coalesced,omarImbue}. The CoTM is designed to improve accuracy, memory efficiency, and scalability, where it has demonstrated superior performance in both training and inference processes~\cite{glimsdal2021coalesced}. The inference operations in CoTMs are based on the principles of learning patterns through simple propositional logical expressions, leading to low complexity and interpretability. Additionally, CoTM performs robustly with imbalanced training data and supports sparse classification, positioning it as a viable alternative to traditional arithmetic-driven ML algorithms like deep neural networks (DNNs). In this paper, we make the following four major contributions:
\begin{itemize}
    \item[(I)] The development of the first IMC architecture, the IMPACT, which leverages Y-Flash arrays for CoTM inference (Section~\ref{Sec:Performance and Efficiency Evaluation of the IMPACT}, Figure~\ref{fig_CoTM_Arch}). 
     \item[(ii)] Exploring the implementation of the CoTM algorithm's data processing procedure into a hardware framework highlights its scalability (Section~\ref{Sec:CoTM Inference Architecture: The IMPACT}).
    \item[(iii)] Analyzing the resilience of Y-Flash technology to variability, highlighting its potential for real-time ML applications that demand reliable inference (Section~\ref{Sec:Experimental Results}\ref{Sec:Y-Flash Crossbar Variability Analysis}).
    \item[(iv)] Combining Y-Flash devices in IMC structure accelerates inference while ensuring the model's interpretability, which is crucial for domains requiring transparency in decision-making (Section~\ref{Sec:Experimental Results}\ref{Sec:Encoding TAs/Weights}).
    
\end{itemize}
The rest of the paper is structured as follows. Section 2 provides an overview of the CoTM hierarchical layers, inference process, training process, and Y-Flash device characteristics. Section 3 explains the mapping process for Tsetlin Automata and weight matrices within the CoTM framework. Section 4 offers an analysis of device variations and inference accuracy metrics. Section 5 summarizes the paper's key findings and conclusions.

\section{Methods}
\label{Sec:Methods}
This section introduces the two main principles for constructing the proposed architecture: IMPACT. Firstly, CoTM is an algorithm for classification tasks designed to improve efficiency and scalability through weighted clauses. Secondly, the Y-Flash device is used to create the crossbar array, which is known for its multi-conductance capability.

\vspace*{-5pt}

\subsection{CoTM algorithm}
\label{CoTM algorithm}The CoTM is an advancement of the traditional Tsetlin machine (TM)~\cite{2018-granmo-tsetlin,Redress}, an algorithm for machine learning based on propositional logic. In the TM, basic decision-makers called Tsetlin Automata (TA) learn to make binary decisions by receiving rewards and penalties during training. These TAs combine to form clauses. Each class has the same number of clauses that individually vote to either support or oppose its class. This structured approach allows TM to identify patterns in the input data. The CoTM improves the traditional TM by merging "coalescing" clauses. This approach allows the clauses to share TAs, thereby improving overall performance by reducing complexity and expediting the learning process with less computational overhead, as fewer resources are required to maintain and update the TAs. Below, we provide an overview of the algorithmic pipeline, including data preparation and TA states dynamics to form clauses. Following this, the clauses' logical expressions are shared across multiple classes, and finally, how these coalesced clauses are used collectively to classify input data into different classes. 

\textbf{Data preparation:} is a crucial initial step in the CoTM framework. This process involves transforming raw data $(X_i)$ into a structured format suitable for CoTM processing. Typically, raw data points consist of numerical values representing various features. These features are encoded into a binary format. Thresholds and determining the number of bits required for each feature based on design specifications and the feature's range. Each bit is considered alongside its negated form, converting the data into Boolean literals $(L)$. The original version of the bit indicates the presence of a feature, while the negated version indicates its absence. This approach ensures that the data is sparse and helps the CoTM accurately identify and learn patterns, leading to more reliable predictions (see Figure~\ref{fig_CoTM_Algo}b). 

\textbf{Tsetlin Automata:} The CoTM employs a specific type of finite state machine known as the TA as shown in Figure~\ref{fig_CoTM_Algo}a. TA is the learning element of CoTM and is responsible for local decision-making. Each TA consists of 2N states, divided equally into two actions: Include (logic ``1'')  and Exclude (logic ``0''). During training, the TA adapts its state through reinforcement learning principles. State adjustment can be moved toward one of the edges if receiving rewards or move toward the boundary between the two actions if receiving a penalty. Each TA corresponds to a Boolean literal. The interaction between the TA and the corresponding Boolean literal involves two propositional logic gates. The TA's action is inverted through a NOT gate and then interacts with the Boolean literal in an OR gate. Once the training is complete and convergence is reached, the half where the TA's final state is located determines whether the related Boolean literal should be included or excluded in the logical expression, thereby contributing to the classification decision. This mechanism ensures that the CoTM can efficiently learn and represent patterns within the data (see Figure~\ref{fig_CoTM_Algo}b).

\textbf{Clauses:}\label{section:CoTM_Clauses} are logical expressions derived from the decisions made by multiple TAs toward the Boolean literals (see Figure~\ref{fig_CoTM_Algo}b). Each clause is formed by $K$  of TAs participating in the clause's logical expression. A clause can be seen as a rule that evaluates the presence or absence of specific features in the final classification decision. Each clause effectively votes either in favor of (outputting ``1'') or against (outputting ``0'') the class. The collective decisions of all clauses determine the overall output by voting on the under-classification class. Mathematically, a clause $C_j$ can be represented as:
\[ C_j = \bigwedge_{i=1}^K (L_i \lor \sim TA_i) \]
Here, $j$ can be $\{1, \ldots, n\}$, where $n$ is the number of clauses in the CoTM, and it depends on the size of the classification problem. $K$ is the number of Boolean literals. Therefore, the clause matrix size is $(K \times n)$.

\textbf{Weight Matrix} is a two-dimensional array $(W)$ in which each row represents a class, and each column represents a clause (see Figure~\ref{fig_CoTM_Algo}c). The element in the matrix represents the weight of the associated clause's importance in predicting the corresponding class. During training, the clause weights are stochastically adjusted based on their success in predicting the class. Clauses that contribute effectively to accurate predictions may have their weights increased, while those leading to errors may have their weights decreased. This stochastic weight adjustment enables the CoTM to adapt to the data's characteristics and optimize its performance over time. Mathematically, the weight matrix $W$ can be represented as:
\[
W = \begin{bmatrix}
    w_{11} & w_{21} & \cdots & w_{1n} \\
    w_{21} & w_{22} & \cdots & w_{2n} \\
    \vdots & \vdots & \ddots & \vdots \\
    w_{m1} & w_{2m} & \cdots & w_{mn}
\end{bmatrix}
\]

Where:
\begin{itemize}
    \item $w_{ij}$ represents the weight associated with the contribution of clause $j$ to class $i$.
    \item $i$ varies from 1 to $m$ (number of classes).
    \item $j$ varies from 1 to $n$ (number of clauses).
\end{itemize}
\begin{figure}[t]
\centering\includegraphics[width=\linewidth]{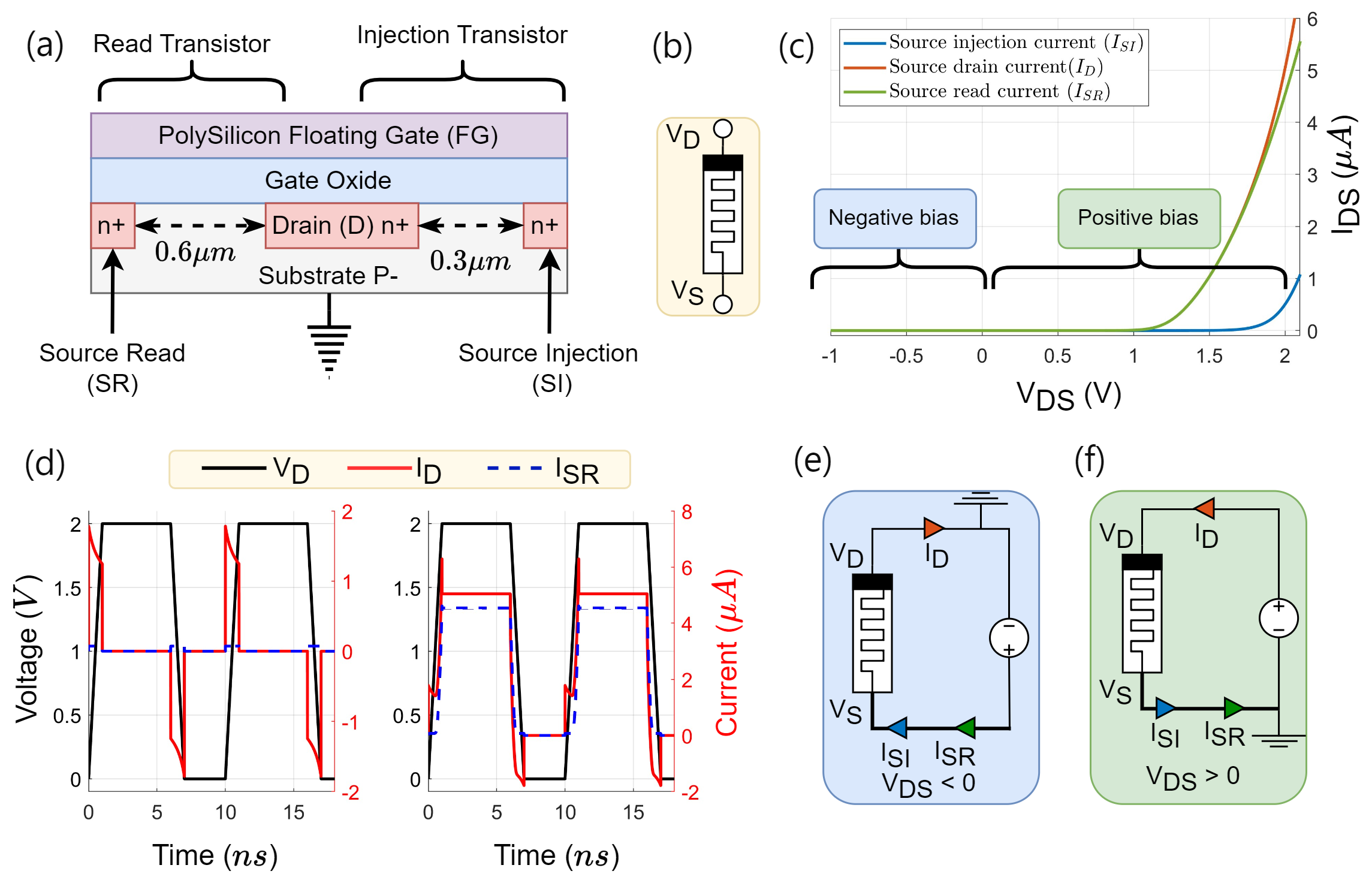}
\caption{(a) The structure of a Y-Flash cell. (b) The symbol of two terminal Y-Flash devices. (c) The Y-Flash characteristics under negative bias $(V_{DS} < 0\ V)$ and positive bias $(V_{DS} > 0\ V)$ reading cycle where (e) and (f) are the two biasing configurations. (d) The simulated DC behavior for $(I_D - V_D)$ and $(I_{SR} - V_D)$ characteristics of the Y-Flash device during a reading pulse $(V_R = 2\ V)$ showing the two distinct Boolean conductance states HCS and LCS.}
\label{fig_Yflash_struct}
\end{figure}

\textbf{Class Computation} includes collecting clause contributions weighted by importance. Each clause produces a vote for a class based on its weight and evaluation result. The weighted sum of these votes determines the predicted class. The output vector \((Y)\) is determined by a majority vote, where each class is assigned a value of 1 if its weighted sum \((V)\) is positive and 0 if its weighted sum is negative. This method ensures that the final output accurately reflects the majority decision of the weighted class contributions.
\[
V = W \cdot C \quad \text{and} \quad Y_i = \begin{cases}
    1, & \text{if } V_i > 0 \\
    0, & \text{if } V_i \leq 0
\end{cases}
\]
\begin{itemize}
    \item $V$ represents the signed weighted sum vector.
    \item $Y_i$ represents the value of class i in the output Boolean vector Y.
\end{itemize}
\textbf{Training} iteratively adjusts the states of the TAs and the weights of the clauses based on the input Boolean literals and the expected output. This process applies reinforcement learning, where correct classifications reinforce the current state and incorrect classifications make fitting adjustments. The objective is to minimize classification errors by refining the clauses' logical expressions formed by the TAs and optimizing the weight matrix.


\begin{table*}[t]
\caption{Y-Flash's operational modes for three- and two-terminal configurations~\cite{YFLASHMODEL_physical}.}
\label{table:Y-FLASH's operation modes}
\centering
\begin{tabular}{lllll}
\hline
\textbf{Operation} & \multicolumn{4}{c}{\textbf{Terminals configurations}} \\
\cline{2-5}
\textbf{Mode} & \textbf{Drain} & \textbf{Source Read} & \textbf{Source Injection} & \textbf{Substrate} \\
\hline

\multirow{2}{*}{Read} & $V_{R}=2V$ & GND & GND & GND \\

 & $V_{R}=2V$ & GND & Z & GND \\ \hline

\multirow{3}{*}{Program} &$V_{P}=5V$ &GND & GND & GND \\
 & $V_{P}=5V$ & Z & GND & GND \\
 & $V_{P}=5V$ & Z & Z & GND \\ \hline
\multirow{3}{*}{Erase} & GND / Z & $V_{E}=8V$ & $V_{E}=8V$ & GND \\
 & GND / Z & GND / Z & $V_{E}=8V$ & GND \\
 & (1 < $V_{Des}$ < 2V) & GND / Z & $V_{E}=8V$ & GND \\ \hline
\multicolumn{5}{l}{$Z$: is high impedance or floating, $V_{Des}$ Deselect mode.} \\
\end{tabular}
\vspace*{-4pt}
\end{table*}

\subsection{Y-Flash Devices}
\label{section:Y-Flash Devices}
Y-Flash technology bridges the gap between memristors and standard CMOS technology. It was fabricated using a 180 nm CMOS process~\cite{YFLASH2022_DBNN}. In the structural diagram of a single Y-Flash device, as shown in Figure~\ref{fig_Yflash_struct}a, two transistors, the read transistor (SR) and the injection transistor (SI), are connected in parallel. Both transistors share a standard drain (D) terminal and a polysilicon floating gate (FG). The SR is designed with a more extended channel $0.6\ \mu m$ to achieve a lower threshold voltage $\approx0.3\ V$, facilitating the reading operation with a low voltage. The SI has a shorter channel length $0.3\ \mu m$ and a higher threshold voltage $\approx1.5\ V$ to enable high-voltage for programming and erasing cycles ~\cite{YFLASH2019_2terminal}. This design ensures that the Y-Flash device can effectively operate across a range of reading voltages, making it an adaptable component for IMC applications. Table~\ref{table:Y-FLASH's operation modes}  outlines the various operational modes of the device. When the Sources of SR and SI are externally connected, this configuration enables the device to operate as a two-terminal memristor as well as simplicity in the architecture array design, as shown in the related symbol Figure~\ref{fig_Yflash_struct}b. All data and measurements presented in this paper were performed utilizing the two-terminal mode configuration. 
\begin{figure}[t]
\centering\includegraphics[width=\linewidth]{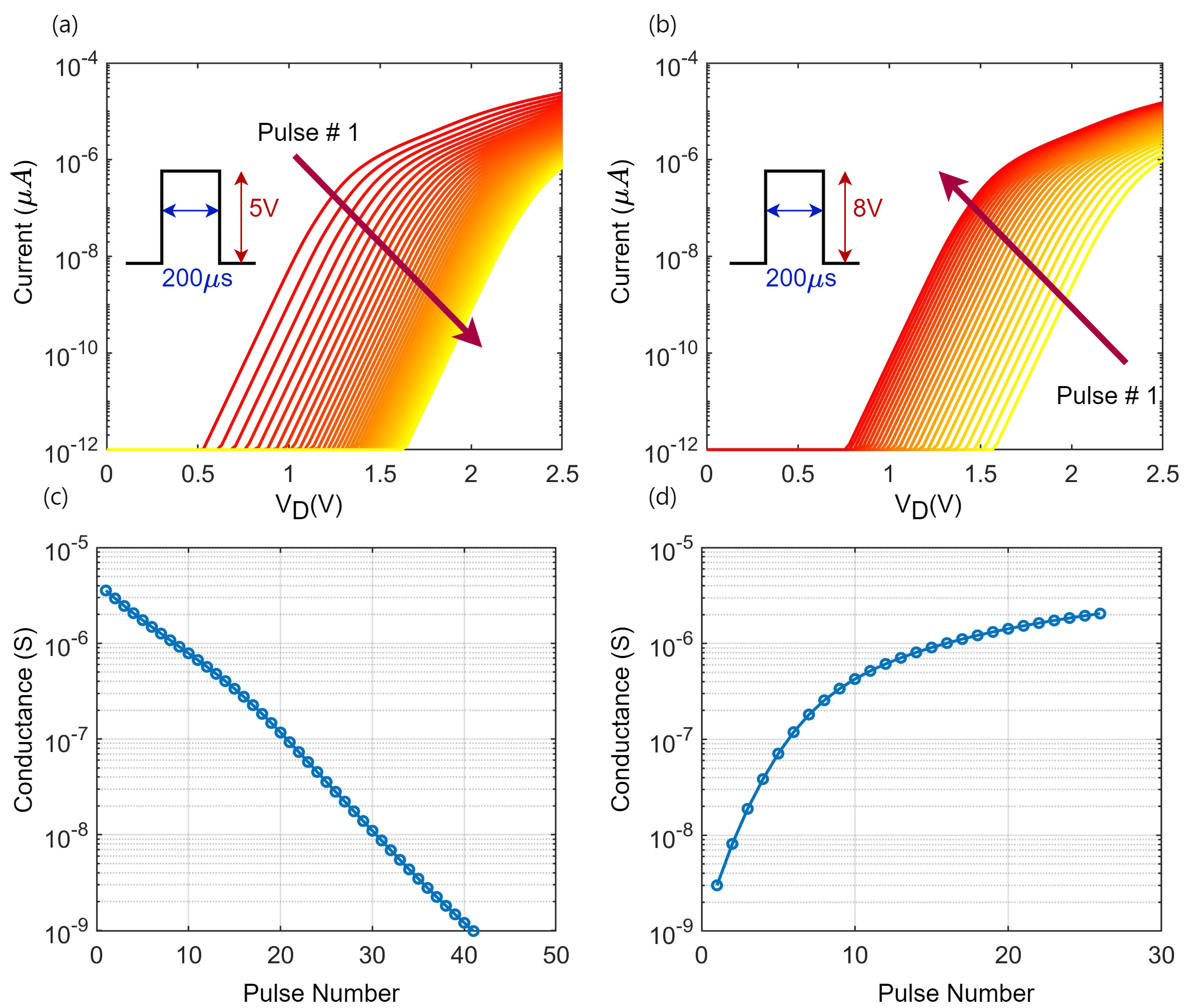}
\caption{The simulated analog turnability behavior of the Y-Flash device. (a) Programming from HCS to LCS. (b) Erasing from LCS to HCS. (c) Conductance values were measured at  $(V_R=2V)$ for the programming cycles in (a). (d) Conductance values were measured at  $(V_R=2V)$ for the erasing cycles in (b).}
\label{fig_Yflash_multi_States}
\end{figure}
A voltage DC sweeping was performed to analyze the device's behavior in both directions. The forward and backward biasing measurements were taken separately and then combined in Figure~\ref{fig_Yflash_struct}c. Figure~\ref{fig_Yflash_struct}f illustrates the configuration of the forward biasing where $V_D > 0$ V and $V_S = 0$ V. The device showed a nonlinear relationship between $I_{DS}$ and $V_{DS}$. Conversely, when the voltage bias is reversed, as the connection shown in Figure~\ref{fig_Yflash_struct}e where $V_D = 0$ V and $V_S > 0$ V, the transistors are switching off where the current flow is negligible due to the floating gate being coupled to the drain. This unique $(I_{DS} - V_{DS})$ behavior allows the device to be self-selected and minimize sneak-path currents, as well as eliminating the need for the selector device when using the device in an array crossbar. Figure~\ref{fig_Yflash_struct} displays the reading pulses of the device operating in Boolean mode. The device was subjected to a sequence of program pulses $(V_p=5\ V \text{and width =}200\ \mu s)$, followed by a reading pulse $(V_R=2\ V \text{and width =}5\ ns)$  applied to the drain until reaching the $LCS \approx 1\ nS$. Under this condition, the output read current was $I_{SR} \approx 1\ nA$, as displayed in linear scales. The process was then repeated to achieve the $HCS$ by applying a series of erase pulses $(V_E=8\ V \text{and width =}200\ \mu s)$, followed by a read pulse. The readings were taken at $HCS \approx 2.2\ \mu S$ with a read current $I_{SR} \approx 4.5\ \mu A$. The overshoot, observed at the rising and falling edges of the reading current pulse measurements, is attributed to the device's internal parasitic capacitors. Figure~\ref{fig_Yflash_multi_States} illustrates the analog turnability behavior of the Y-Flash device, verifying the implementation of multiple states during programming $(V_p=5\ V \text{and width =}200\ \mu s)$ from HCS to LCS and the reverse operation during erasing $(V_E=8\ V \text{and width =}200\ \mu s)$. The $(I_{DS} - V_{DS})$ and conductance shift during the programming phase are shown in Figure~\ref{fig_Yflash_multi_States}a. The programming involves applying voltage cycles starting from HCS until reaching $LCS < 1\ nS$, which increases the charge in the floating gate and decreases the device conductance. Figure~\ref{fig_Yflash_multi_States}b presents the $(I_{DS} - V_{DS})$ and conductance dynamics during the erasing phase. The erasing operation starts from (LCS), which gradually increases the device conductance back until reaching $HCS > 2.2\ \mu S$. The $(I_{DS} - V_{DS})$ characteristics indicate a nonlinear relationship, implying the device's ability to achieve multiple conductance states through controlled programming/erasing cycles.

The combined measured conductance values were taken at $V_R=2\ V$ from the programming and erasing cycles shown in Figure~\ref{fig_Yflash_multi_States}c, and d, respectively. The logarithmic scale used in both subfigures emphasizes the wide range of conductance values the Y-Flash device can achieve, demonstrating its potential for multi-state memory applications. Additionally, when operating under an identical pulse width, the erase process requires fewer cycles to reach the HCS, whereas the programming process requires a longer duration to achieve the LCS. A shorter erase pulse width is necessary for more precise control and accurate restoration to the previous conductance state.

\begin{figure}[!t]
\centering\includegraphics[width=\linewidth]{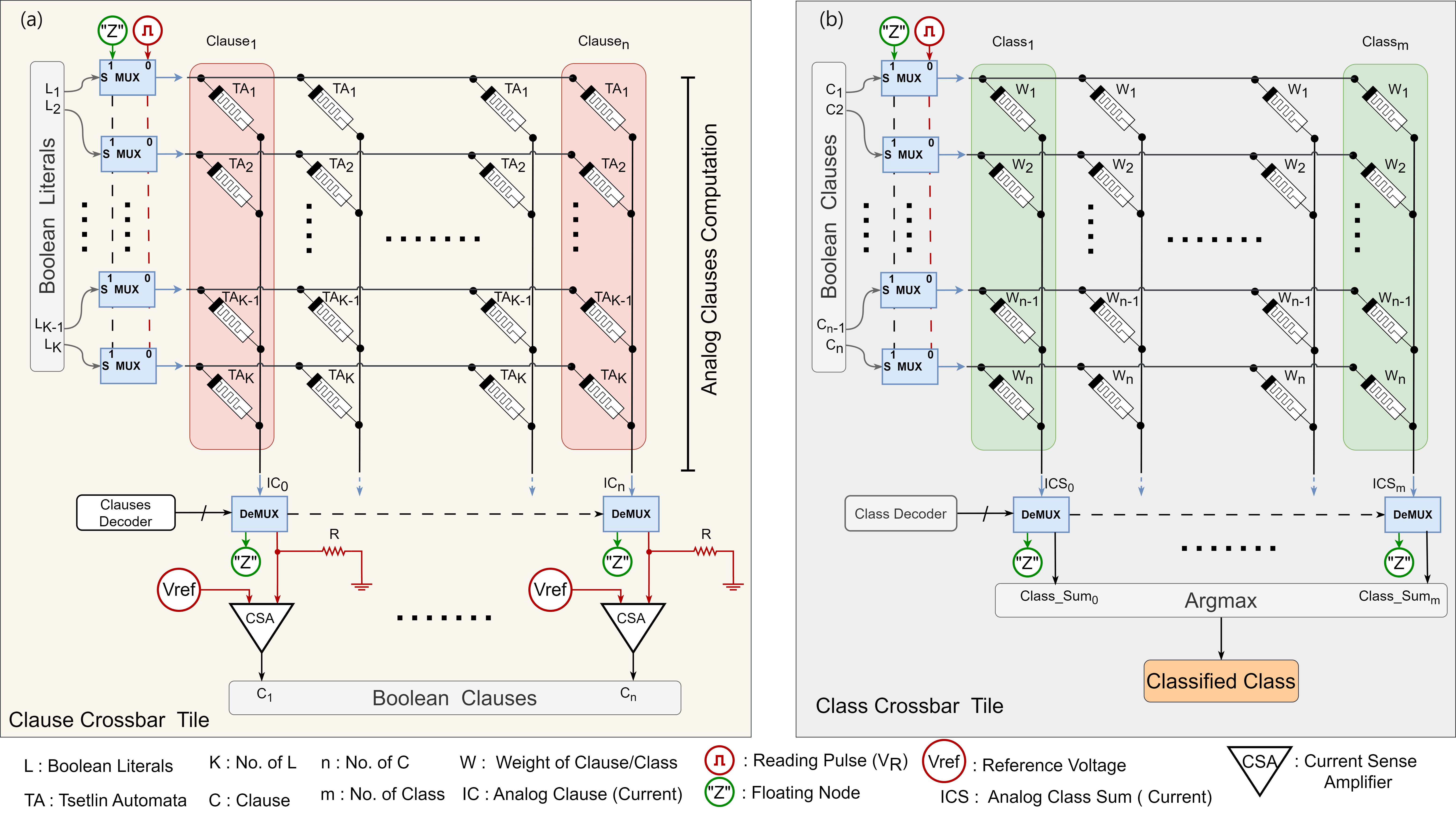}
\caption{The IMPACT architecture shows the two designed arrays. (a) The clause crossbar tile functions in Boolean conductance mode for clause computation. (b) Class crossbar tile works in analog tunable conductance mode for class computation.}
\label{fig_CoTM_Arch}
\end{figure}
\vspace*{-5pt}

\section{IMPACT Architecture} \label{Sec:CoTM Inference Architecture: The IMPACT}

This section presents the IMPACT architecture, which is based on the Y-Flash memristor crossbar. The overall architecture is shown in Figure~\ref{fig_CoTM_Arch}. The IMPACT is designed to perform in-memory computing inference for the CoTM algorithm. To verify the robustness and reliability of IMPACT, two crossbars were assembled and tested using the MNIST dataset: the clause crossbar tile (see Figure~\ref{fig_CoTM_Arch}a) and the class crossbar tile (see Figure~\ref{fig_CoTM_Arch}b). The MNIST dataset consists of \((28\times28)\) grayscale images of the 10 handwritten digits. This results in   \((K = 2 \times 28 \times 28)\) Boolean literals and 10 classes. The clause crossbar tile should have rows equivalent to the number of Boolean literals  \(K\) and columns equal to the number of clauses specified in the design to be 500. To ensure a robust and adaptable architecture, the size of the clause crossbar tile was set at \( (2048 \times 500)\). The size of the class crossbar tile is determined by the number of clauses (rows) and classes (columns), so it is designed to be \((500 \times 10)\). The principles and operations of each tile are examined and supported with experimental measurements in the following subsections.

\subsection{Clause Crossbar Tile}
\label{Clause Crossbar Tile}
The TA actions are mapped into the clause crossbar tile of the trained CoTM on the MNIST dataset as shown in Figure~\ref{fig_CoTM_Arch}a. The array in this crossbar operates in Boolean mode by performing program/erase to an LCS $<1\  nS$ or an HCS $>2.4\ \mu S$. Each TA's action is mapped to a single Y-Flash cell in the array, according to the mapping mechanism outlined in Table~\ref{Table:Ta mapping}. When the TA's action is inclusion, indicated by logic ``1'' it is translated to an HCS in the corresponding Y-Flash cell. Conversely, when the action is exclusion, indicated by logic ``0'' it is translated to an LCS in the related cell in the crossbar array. The input Boolean literals are converted to voltage levels in such a way that the Boolean literal ``1'' and literal ``0'' are applied to the crossbar as a floating node ``Z'' or a reading voltage \(V_R = 2\ V\), respectively. This logic representation simulates the digital inverter gate behavior after the TA in the CoTM algorithm. Leveraging the multiplication in Ohm's law at the crosspoint between the programmed TA and the input literals mimics the OR gate in the algorithm, where the interaction between the TA action and the input literals is executed. This can be expressed as follows
\[I = TA \times L \times V_R\]
\begin{itemize}
    \item $I \in (\approx 0\, ,\, \approx 5\mu)A$  is the interaction output current. 
    \item $TA \in (<\, 1n\, , \, >\, 2.4\mu)S$.
    \item $L \in (0\,, 1)$ used as a selector for the row multiplexer.
    \item $V_R = 2V$.
\end{itemize}
 This approach ensures that the CoTM digital logic operations are faithfully replicated in the clause crossbar tile, supporting accurate in-memory computation for the CoTM algorithm.

\begin{figure}[t]
\centering\includegraphics[width=\linewidth]{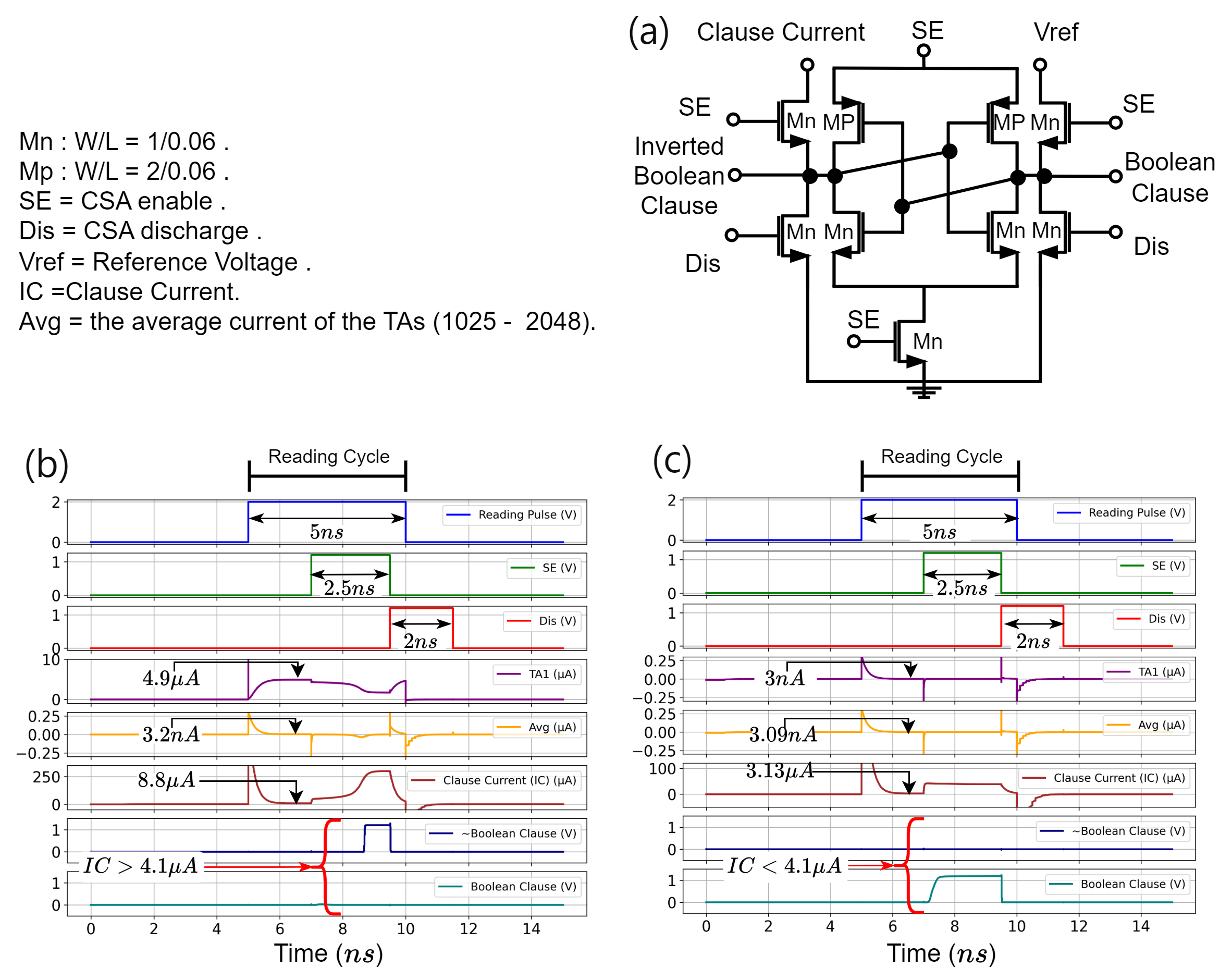}
\caption{Generating the Boolean Clause. (a) Current sense amplifier design. (b) Detecting the include of literal ``0''  during the reading cycle. Where \(TA_1\) in the clause is storing an include action and interacted with input literal ``0''. TAs \((2\ to\ 1024)\) interacted with literal ``1'', and TAs \((1025\ to\ 2048)\) storing exclude actions and all interacted with input literal ``0''. (c) Testing the worst-case scenario during the reading cycle. Where all the TAs in a clause storing exclude actions, half of them received input literals ``0'' while the other half received input literals ``1''.}
\label{fig_CSA}
\end{figure}

The CoTM algorithm performs clause computation using the logic AND gate (see Figure~\ref{fig_CoTM_Algo}b). The clause output is assigned a logic ``0''  \textit{if at least one literal ``0'' interacted with an include action}, regardless of the other TAs and their interacting literals. This principle is implemented in the IMPACT architecture through the clause computation tile. Each clause is defined by one column integrated with a Current Sense Amplifier (CSA) at the end of the column. Accordingly, the clause computation is governed by Kirchhoff's current law, which accumulates all interaction currents produced by the TAs associated with the clause during the reading cycle. Then the clause voltage is the potential drop across a resistor connected to the input of the CSA. Figure~\ref{fig_CSA}a displays the design of the CSA. It has two input terminals connected to the clause voltage and a reference voltage (Vref). Vref determines the voltage range that the input terminals of the CSA can tolerate while accurately sensing the clause state. The CSA has two outputs: the Boolean clause and its inverted version. A simple latch serves as the sensing element. The Sense Enable (SE) activates the CSA in the reading cycle and charges one of the outputs to the SE level, while the second output will be near the common voltage. SE also helps to reduce leakage power. The Discharge signal (Dis) balances the two CSA outputs by discharging them to a common voltage level, preparing the CSA for the next reading cycle.
\begin{table*}[t]
\caption{Translating the TA's actions and the Boolean literals to conductance and voltage, respectively.}
\label{Table:Ta mapping}
\centering
\begin{tabular}{lllll}
\hline
\textbf{Input} & \textbf{Applied} & \textbf{TA}& \textbf{Programmed} &\textbf{Interaction}\\

\textbf{literals} & \textbf{voltage(V)} & \textbf{action} & \textbf{conductance($\mu S$)} & \textbf{current ($\mu A$)} \\
\hline

\multirow{2}{*}{Logic ``1''} &  ``Z''  & Include & $> 2.4 $ & $\approx$ 0 \\
  & ``Z'' & Exclude & $< 1 \times 10^{-3}$  & $\approx$ 0 \\ \hline
  \cline{2-5}
\multirow{2}{*}{Logic ``0''} & \cellcolor{cyan!20}$V_R =2$ & \cellcolor{cyan!20}Include & \cellcolor{cyan!20}$> 2.4 $ &\cellcolor{cyan!20} $\approx$ 5 \\
  & $V_R =2$ & Exclude & $< 1 \times 10^{-3}$ & $\approx$ $1 \times 10^{-3}$\\ \hline
\end{tabular}
\vspace*{-4pt}
\end{table*}
The CSA operational boundaries are designed to differentiate between the Boolean states of the clause when evaluating the clause analog voltage. These two boundaries are set based on the data in Table~\ref{Table:Ta mapping} and CoTM inference mechanism.
According to Table~\ref{Table:Ta mapping}, if at least one literal ``0'' interacts with TA that stores an include action, the produced output current will be \(\approx5\ \mu A\), which is sufficient to set its Boolean clause output to ``0'', otherwise it will be ``1''. This specifies the first boundary for the CSA's operation. The second boundary, describing the worst-case scenario, occurs when literal ``0'' interacts with TA that stores an exclude action, resulting in a current of \(\approx1\ nA\). It is crucial to ensure that this current accumulation in the clause does not reach the level of the first boundary. If many of these currents accumulate in the clause, there is a risk that the CSA will produce a Boolean clause with ``0'' and introduce an error in the classification process. Furthermore, in the CoTM algorithm, all inputs are the negated version of the first half of the input binary features. Considering our design array size, the critical point for testing is whether there will be 1024 TAs storing exclude actions and literal ``0'' applied to all of them.

Figure~\ref{fig_CSA} shows the successful implementation of these conditions, even with the size of our array. In Figure~$\ref{fig_CSA}$b, the first boundary conditions detect the include of the literal ``0'' during the reading cycle. In this scenario, \(TA_1\) in the clause stored an include action (HCS) and interacted with input literal ``0'' \((V_R=2\ V)\). TAs (2 to 1024) interacted with literal ``1'' (``Z''), and TAs (1025 to 2048) stored exclude actions (LCS), and all interacted with input literal ``0'' \((V_R=2\ V)\). Based on these conditions and as seen in Figure~\ref{fig_CSA}a, the computed Boolean clause is ``0'', which matches the CoTM algorithm. The second boundary of CSA tests the worst-case scenario during the reading cycle, where all the TAs in a clause stored exclude actions (LCS), with half of them receiving input literal  ``0'' \((V_R=2\ V)\) and the other half receiving input literals ``1'' (``Z''), see Figure~\ref{fig_CSA}c. Under these conditions, each TA will produce an average current \(\approx 3\ nA\). This current is higher than the expected \(\approx 1\ nA\) due to the nonlinearity in the device behavior. However, the computed clause output current is \(\approx 3.1\ \mu A\). This current translates to a Boolean clause ``1'' and matches the CoTM algorithm. The clause computation is completed within \(5\ ns\), using three sequential pulses: an initial reading pulse of \(5\ ns\), within which a \(2\ ns\) delay followed by a \(2.5\ ns\) SE pulse to activate the CSA. This delay is essential for overcoming the effect of parasitics and stabilizing the output of the current clause. After the SE is completed, a Dis pulse is applied within the last \(500\ ps\) of the reading pulse to reset the CSA and prepare it for the next reading cycle. The CSA design is robust and effectively mirrors the CoTM functional requirements. A clause current of \(4.1\ \mu A\) or higher translates to a Boolean ``0'', excluding the clause from voting in the current classification cycle; otherwise, the Boolean clause will be ``1''. Using CSA and Kirchhoff's current law, clause computations were mapped from the analog domain to the Boolean domain. The clause computation can be expressed in mathematical expression as follows:
\[
IC_j =  \sum_{i=1}^{K} \left( TA_{ji}\ L_i\ VR \right)
\]

\[
C_j = \begin{cases}
    1, & \text{if } IC_j < 4.1\ \mu A \\
    0, & \text{if } IC_j > 4.1\ \mu A
\end{cases}
\]
Where
\begin{itemize}
    \item $IC$: The analog clause (current).
    \item $C$: The Boolean clause \(\in(0,1)\).
    \item $j$: The $j$-th clause \{1,...,n\}.
    \item $i$: The $i$-th literal \{1,...,K\}.
\end{itemize}

\begin{table}
\caption{CSA Corner and process variation analysis using VDD 1.2V: All voltage readings are in mV.}
\label{table:CSA_analysis}
\centering
\begin{tblr}{
  width = \linewidth,
  colspec = {Q[70]Q[300]Q[70]Q[54]Q[54]Q[54]Q[54]Q[54]Q[54]Q[110]},
  row{1} = {c},
  cell{1}{1} = {r=2}{},
  cell{1}{2} = {r=2}{},
  cell{1}{4} = {c=5}{0.297\linewidth},
  cell{1}{9} = {c=2}{0.137\linewidth},
  cell{2}{4} = {c},
  cell{2}{5} = {c},
  cell{2}{6} = {c},
  cell{2}{7} = {c},
  cell{2}{8} = {c},
  cell{2}{9} = {c},
  cell{2}{10} = {c},
  cell{3}{1} = {r=4}{c},
  cell{3}{2} = {r=2}{c},
  cell{5}{2} = {r=2}{c},
  cell{7}{1} = {r=4}{c},
  cell{7}{2} = {r=2}{c},
  cell{9}{2} = {r=2}{c},
  cell{11}{1} = {c=10}{0.931\linewidth},
  hline{1,3,7,11} = {-}{},
  hline{2} = {3-10}{},
}
Reading & Operation & CSA & Corner &  &  &  &  & ~Process
  Variation & \\
 &  & Output & T~ & FF & SS & SF & FS & Mean & SD\\
\begin{sideways}Before the inverter\end{sideways} & The clause is "0" if the ("0", Include) pair\(^*\) is present in the column & C\(^{**}\) & 865.7 & 860.3 & 867.1 & 892 & 819.6 & 865.3 & 7.334\\
 &  & \(\sim\) C & 0.951 & 3.956 & 0.317 & 1.324 & 0.999 & 1.008 & 0.32\\
 & The clause is "1" if the ("0", Include) pair is absent in the column & C & 28.35 & 30.06 & 28.33 & 28.75 & 28.38 & 28.4 & 0.158\\
 &  & \(\sim\) C & 876.4 & 872.9 & 874.7 & 908.1 & 823.1 & 875.9 & 8.729\\
\begin{sideways}After the inverter\end{sideways} & The clause is "0" if the ("0", Include) pair is present in the column & C & 0.753 & 4.216 & 0.119 & 1.923 & 0.47 & 0.827 & 0.394\\
 &  & \(\sim\) C & 1200 & 1200 & 1200 & 1200 & 1200 & 1200 & 0.46$\times 10^{-3}$\\
 & The clause is "1" if the ("0", Include) pair is absent in the column & C & 0.577 & 3.311 & 0.096 & 1.302 & 0.431 & 0.632 & 0.296\\
 &  & \(\sim\) C & 1200 & 1200 & 1200 & 1200 & 1200 & 1200 & 0.65$\times 10^{-3}$\\
\(^{**}(\sim\)) C : (Inverted) Boolean Clause, \(^*\) Pair : (Literal, Action) &  &  &  &  &  &  &  &  & 
\end{tblr}
\end{table}

\begin{figure}[t]

\centering\includegraphics[width=\linewidth]{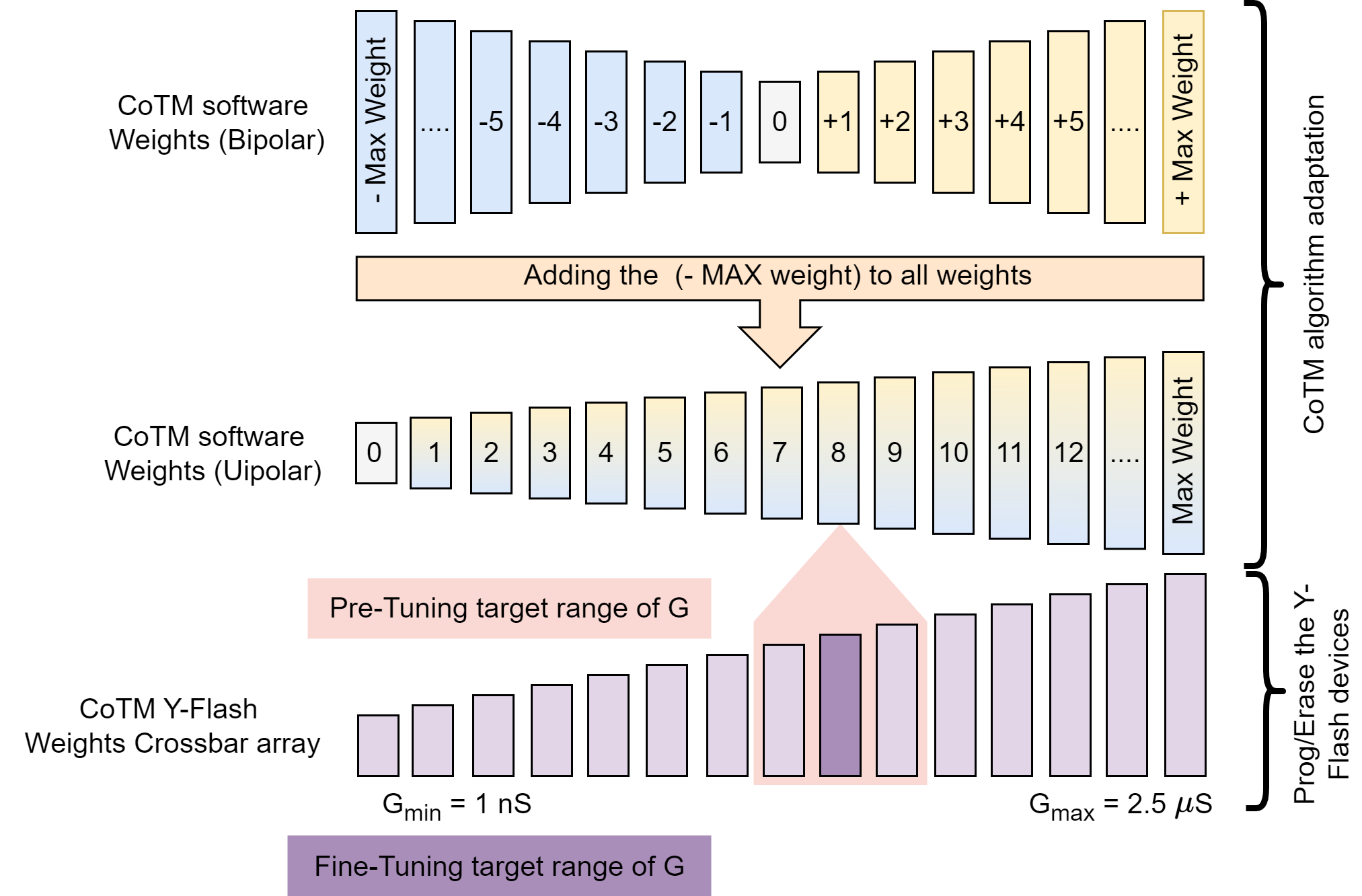}
\caption{Tuning the memristor conductance involves two stages for mapping the CoTM weights into the class crossbar tile. Initially, the integer software weights are adjusted by incrementing them by the absolute value of the smallest weight to ensure they are unsigned, making the minimum weight 0, which will be mapped to LCS. In the pre-tuning stage, the memristor's conductance is programmed to be within the target range using a wider pulse and a more tolerant error range. Following this, the fine-tuning stage further refines the conductance to be correlated with the target conductance by employing a smaller pulse width and a narrower acceptance range.}
\label{fig_Tuning_procedure}
\end{figure}
\vspace*{-5pt}

To assess the impact of process variations and mismatches in the design of the CSA, a corner analysis was conducted, as indicated in Table~$\ref{table:CSA_analysis}$. This investigation examined typical conditions (TT) alongside the corner cases, Fast-Fast (FF), Slow-Slow (SS), Slow-Fast (SF), and Fast-Slow (FS). The voltage readings of the two direct outputs of the CSA, the Boolean clause (C) and the inverted Boolean clause ($\sim$ C) for both before and after the inverter, are provided,  where the inverter amplifies the CSA's output sensed value. The analysis was taken during both the present and the absence of the literal action pair ("0", Include) (the critical cases), as seen in Table~$\ref{table:CSA_analysis}$. The results indicated that the CSA outputs showed small standard deviations (SD) for both CSA outputs under the corner conditions, demonstrating minimal variation. The analysis showed that the output readings before the inverter were within acceptable ranges across different conditions. Moreover, the Monte Carlo simulations performed for over 2000 samples showed nominal mismatch effects on these output voltages, verifying the robustness of the CSA design. The output of the inverters switched from VDD to GND and vice versa, in accordance with the outputs from the C and $\sim$ C. Furthermore, increasing the transistor's width to length (W/L) ratio reduces mismatch by enhancing device area, thus improving circuit stability. Overall, the CSA performs reliably under process variations, ensuring minimal impact on computing time and bit errors in the system performance.

\subsection{Class Crossbar Tile}
\label{Class Crossbar Tile}
The weights of a trained CoTM model for the MNIST dataset are merged into the class crossbar tile, as shown in Figure~\ref{fig_CoTM_Arch}b. The array in this crossbar operates in a tunable analog mode, using varied pulse widths during programming to achieve lower conductance states and during erasing to achieve higher conductance states. Each weight (W) is assigned to a single Y-Flash cell in the array, with higher weights corresponding to higher conductance in the respective Y-Flash cell and lower weights corresponding to lower conductance. As shown in Figure~\ref{fig_Tuning_procedure}, the CoTM produces signed weights (bipolar). Since memristor arrays naturally support unipolar conductance states and avoid handling both positive and negative currents within the same crossbar array, which can be challenging and power-consuming. These weights are converted to unsigned (unipolar) by adding the most negative weight (\(W_{\text{min}}\)) in the trained CoTM model to each weight in the matrix:

\[
W_{ij} = W_{ij} + |W_{\text{min}}|
\]

This modification ensures that the classification process remains unaffected, as the same constant shift is applied uniformly across all weights, preserving their relative differences. Additionally, it maintains the simplicity of the IMPACT architecture by avoiding the need for additional circuits to manage bipolar weights, thus reducing design complexity and potential sources of error.

In order to achieve precise mapping of the weights into the class crossbar tile, a two-step tuning process is utilized: pre-tuning followed by fine-tuning. The entire conductance range of the Y-Flash device (with an equivalent current range of \(1\ nA - 5\ \mu A\)) is divided into segments. The number of these segments corresponds to the highest weight in the trained CoTM model. In the pre-tuning phase, a wider pulse width \(500\ \mu s\) is rapidly used to attain the target conductance associated with the weight while minimizing power consumption by reducing the number of the required program/erase pulses. In this phase, a higher error margin is accepted, allowing the conductance to fall within \(\pm 20\) segments of the target conductance. After pre-tuning, fine-tuning is carried out to adjust the conductance state closer to the target value. In this phase, a narrower pulse width \(50\ \mu s\) is utilized to achieve higher precision, with a reduced error margin of \(\pm 5\) segments. This approach aims to achieve an efficient balance between accuracy and power consumption, ensuring that the conductance values closely match the desired weights while minimizing energy usage.

The inputs are the sampled Boolean clauses (C) from the clauses crossbar tile. These Boolean clauses are converted to voltage levels, where Boolean clause outputs ``1'' and ``0'' are applied to the crossbar as  \(V_R = 2V\) and a ``Z'', respectively. By summing the currents at the crosspoint between the stored weight in the class crossbar tile and the input Boolean clauses, the CoTM algorithm achieves its multiplication behavior. Each multiplication output current represents the weighted vote of that applied input Boolean clause. Kirchhoff's current law is then applied to sum these currents from all the weights (rows) connected to the class (column), functioning as a votes popcount mechanism to produce the class-weighted sum for each class. This ensures that the CoTM model can effectively perform inference by translating the weight mappings and input voltage levels into accurate classification outputs through the physical laws of electronics. This process can be mathematically expressed as follows:
\[
ICS_j = \sum_{i=1}^{n} W_{ji}\  C_{i}\ V_R
\]
Where
\begin{itemize}
    \item $ICS$: The class weighted sum (current).
    \item $C$: The Boolean clause \(\in(0,1)\) is used as a selector for the row multiplexer.
    \item $j$: The \(j^{th}\) class \{1,...,m\}.
\end{itemize}
The final classified class is determined using an argmax in the decision-making process. The classified class is identified as the maximum class-weighted sum, ensuring that the selection is based on the most significant summed votes (currents) in the Class crossbar tiles.

\begin{figure}[t]
\setlength\belowcaptionskip{-1.0\baselineskip}
\centering\includegraphics[width=\linewidth]{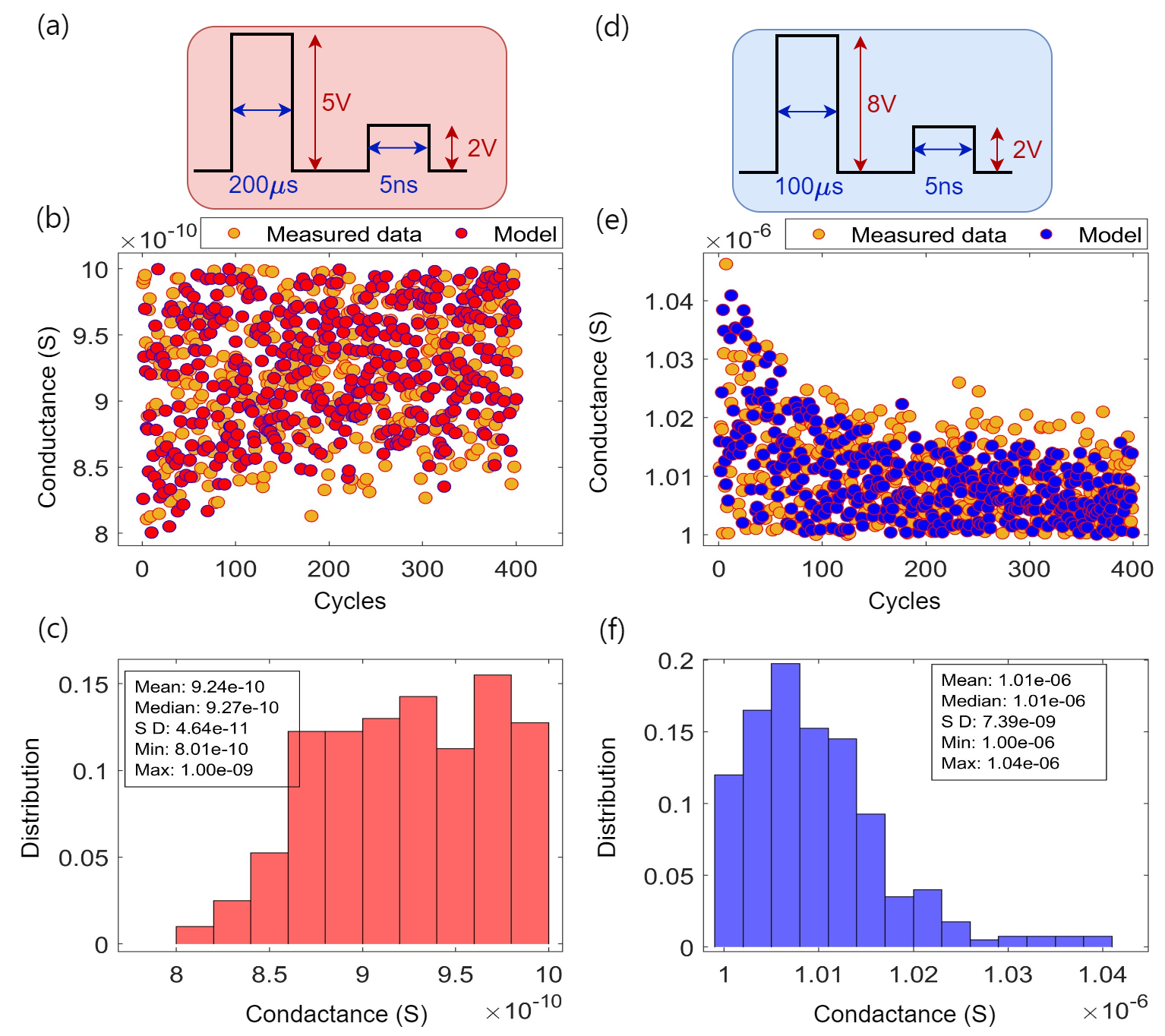}
\caption{An analysis of the cycle-to-cycle variation over 400 cycles. (a) The programming pulse cycle for the transition of the device's state from HCS to LCS $< 1$ nS. (b and c) Display these LCS values and the associated statistical metrics. (d) The erasing pulse cycle for the transition of the device's state from  LCS to HCS $>1\  \mu$ S. (e and f) Provide the distribution of these HCS values and their associated statistical values.}
\label{fig_Y-flash_C2C}
\end{figure}
\vspace*{-5pt}

\begin{figure}[t]
\centering\includegraphics[width=\linewidth]{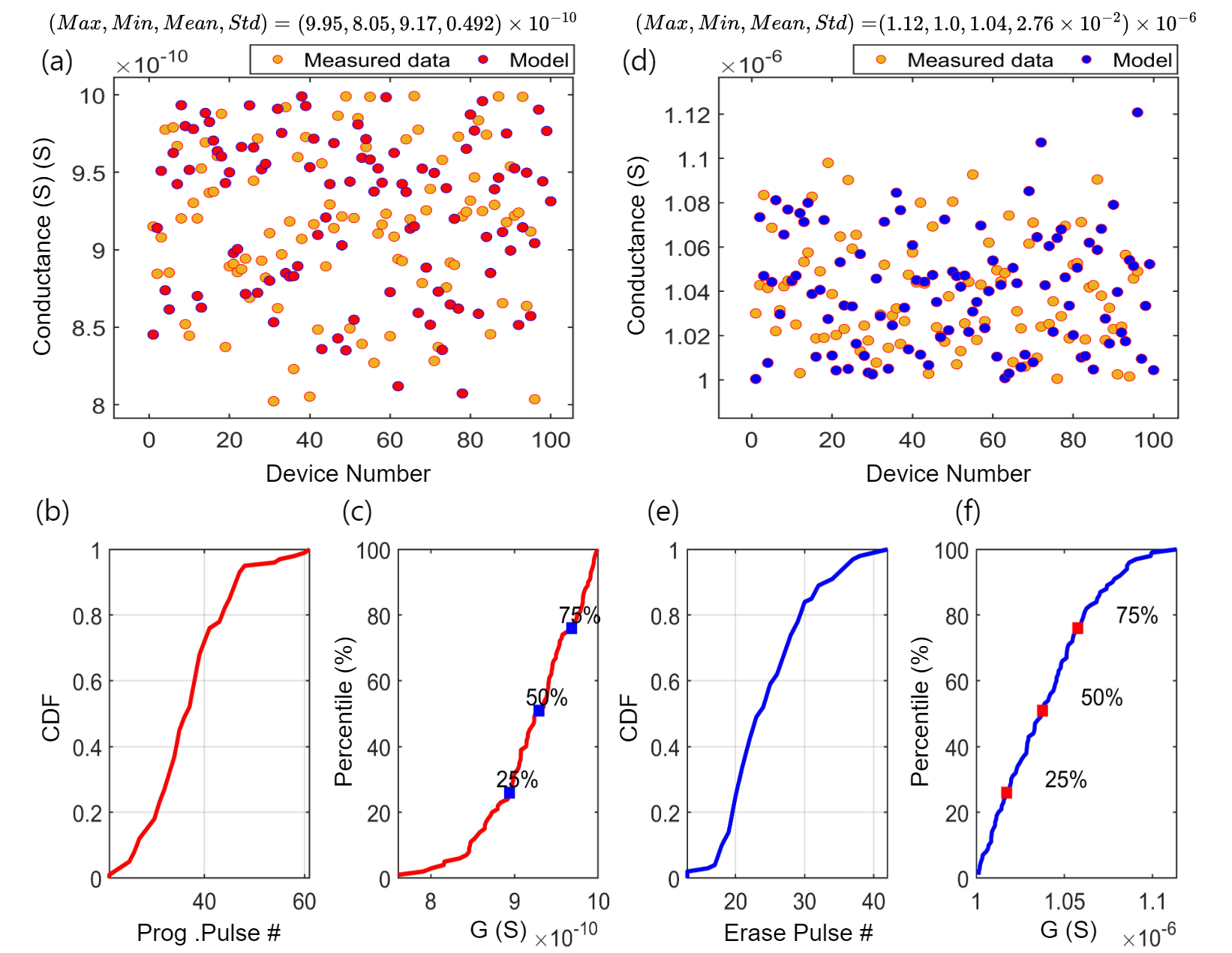}
\caption{The device-to-device variation in the conductance of Y-Flash. (a-c) the programming pulse, the measured and simulated LCS, and the distribution of LCS, respectively. (d-f) the erasing pulse, the measured and simulated HCS, and the
distribution of HCS, respectively.}
\label{fig_Y-flash_D2D}
\end{figure}
\vspace*{-5pt}

\section{Experimental Results}\label{Sec:Experimental Results}
This section explores the resilience of the IMPACT architecture by examining its performance under device-to-device (D2D) and cycle-to-cycle (C2C) variations. It evaluates the reliability of IMPACT in accurately encoding the TAs and weights into the clause and class crossbar tiles.

\subsection{Y-Flash Crossbar Variability Analysis}\label{Sec:Y-Flash Crossbar Variability Analysis}

This section presents the results of simulations and measurements stemming from D2D and C2C variations within the Y-Flash device. The devices were subjected to multiple programming and erasure cycles, and the resulting changes in conductance were recorded and analyzed. Each programming cycle is a full swing from the \(HCS > 1\ \mu S\) until reaching the LCS < 1 nS. Similarly, each erasing cycle is a full swing from  LCS < 1 nS until reaching  HCS > 1 $\mu$S. After each programming or erasing pulse, a reading pulse was applied to record the conductance values. The simulation results were then compared to measured data from a Y-Flash array of 96 devices arranged in a 12x8 configuration~\cite{YFLASH2022_DBNN,YFLASHMODEL_physical}.

The C2C test utilized pulse width for programming (erasing) \(200\ \mu s\) \((100\ \mu s)\). The experiment was conducted using a single device for simulation  ( also one physical device for measurement data \cite{YFLASH2022_DBNN,YFLASHMODEL_physical}). The device was subjected to  400 cycles, with each cycle composing a sequence of program pulses until reaching \(LCS < 1\ nS\), followed by a sequence of erasing pulses until reaching \(HCS > 1\ uS\). The LCS and HCS values for all cycles were then recorded. Figure~\ref{fig_Y-flash_C2C}a, b, and c show the programming pulse, the measured and simulation LCS, and the distribution of the recorded LCS, respectively. The results indicate the device has an LCS distribution with a mean \(0.925\  nS\), and SD  is approximately \(4.41\times  10^{-11} S\), which is about \(4.8 \%\) of the mean value. This implies that the variation in conductance is relatively small, and the device demonstrated stable performance during programming cycles. Additionally, Figure~\ref{fig_Y-flash_C2C}d, e, and f display the erasing pulse, measured and simulated HCS, and the distribution of recorded HCS. The experiment shows that the distribution of HCS has a mean value of \(1.01\ \mu S\), and SD is approximately \(7.42\ nS\), about \(0.74\%\) of the mean value. This further signifies that the conductance variation during erasing cycles is minimal, indicating high device stability.  These minor variations imply that the device exhibits excellent stability and reliability over multiple cycles. The model's close alignment with the measured data further validates the robustness of the simulation in predicting real-world performance. 

\begin{figure}[t]
\centering\includegraphics[width=\linewidth]{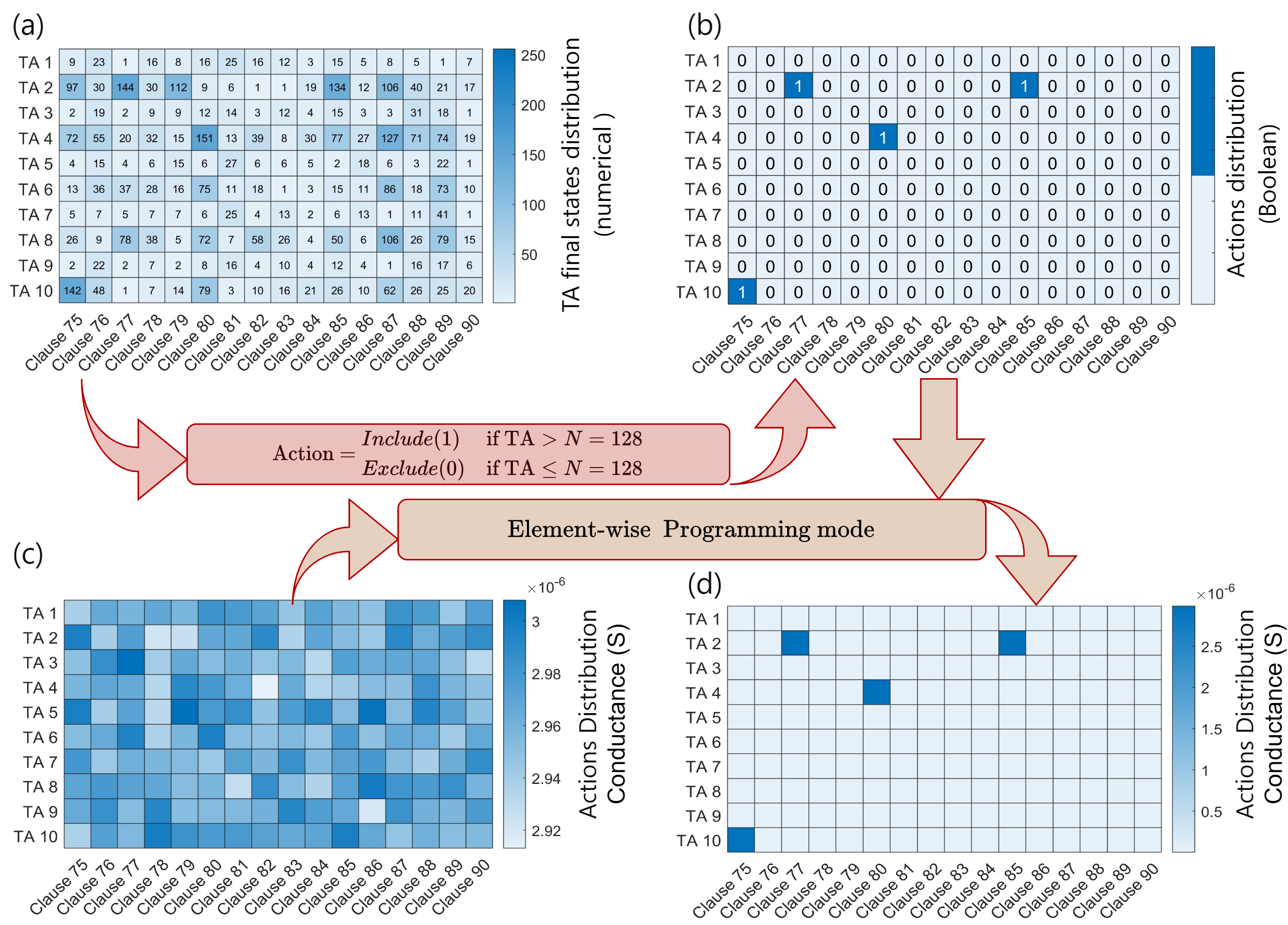}
\caption{Encoding the TA matrix of trained CoTM model for the MNIST dataset on clause crossbar Memory. (a) Represents a sample of the TAs after reaching their final state for the clauses between 75 and 90. (b) Transferring the numerical TA value to one of the two Boolean actions. (c-d) The allocated clause crossbar tile before and after storing the TA action (HCS and LCS). A pulse width of 1 ms is used to program the actions.}
\label{fig_Ta_dist}
\end{figure}

For testing D2D variation, the pulse width of programming and erasing times were set to $200\ \mu s$ and $100\ \mu s$, respectively, for both simulation and measurement experiments. In these experiments, 96 physical devices were used for measurement, while 100 devices were employed for simulations~\cite{YFLASH2022_DBNN,YFLASHMODEL_physical}. Figure~\ref{fig_Y-flash_D2D}a illustrates the devices' recorded LCS. The statistical analysis reveals that the maximum and minimum values are close to the mean \(\approx0.9\  nS\) with an SD of \(0.04\ nS\). This tight clustering around the mean indicates low variability among the devices. Figure~\ref{fig_Y-flash_D2D}b presents the cumulative distribution function (CDF) for the number of programming pulses required per device to reach the recorded LCS values, ranging from approximately 23 to 61 pulses. The sharp rise around pulse number 25 and continued increase until around 48 pulses suggest that most devices required this range of programming pulses, indicating similar programming characteristics. Figure~\ref{fig_Y-flash_D2D}d shows the recorded HCS of the devices, the same as in LCS, with the maximum and minimum values close to the mean \(\approx1.04\ \mu S\) and an SD of \(27.6\ nS\), indicating low variability. As shown in Figure~\ref{fig_Y-flash_D2D}e, the CDF for the number of erasing pulses required per device to reach these recorded HCS values ranges from approximately 15 to 51 pulses. The rapid rise in the middle of the plot between the pulses number 20 to 35 and the flattening near the end indicate that most devices fall within this range, with fewer requiring higher pulse numbers. The percentile plots further support this uniform performance.
Figures~\ref{fig_Y-flash_D2D}c and f display the LCS and HCS percentile plots, respectively. Both plots demonstrate a relatively linear relationship, with the 25th, 50th, and 75th percentiles almost evenly spaced along the pulse axis for the LCS and the HCS. This linearity and narrow spread around the mean suggest a uniform distribution and consistent performance across devices.

The statistical analysis and plots highlight the variability and provide a comprehensive understanding of memristor performance and reliability. The model demonstrated a strong correlation between the simulated and physical devices during the C2C and D2D experiments. The Y-Flash array demonstrates uniform conductance variations in both HCS and LCS states, indicating reliable performance across different devices. Additionally, the Y-Flash's ability to achieve and maintain distinct conductance states using a shorter pulse width makes it suitable for high-density memory storage and complex CoTM implementations. The robustness and minimal conductance variation make these devices ideal for machine learning applications. This ensures dependable weight updates similar to clause weights in a CoTM, resulting in more precise and predictable learning outcomes. Moreover, the relatively uniform programming and erasing processes, as indicated by the CDFs of pulse numbers, further enhance their reliability for repeated training cycles in machine learning algorithms.

\begin{figure}[t]
\centering\includegraphics[width=\linewidth]{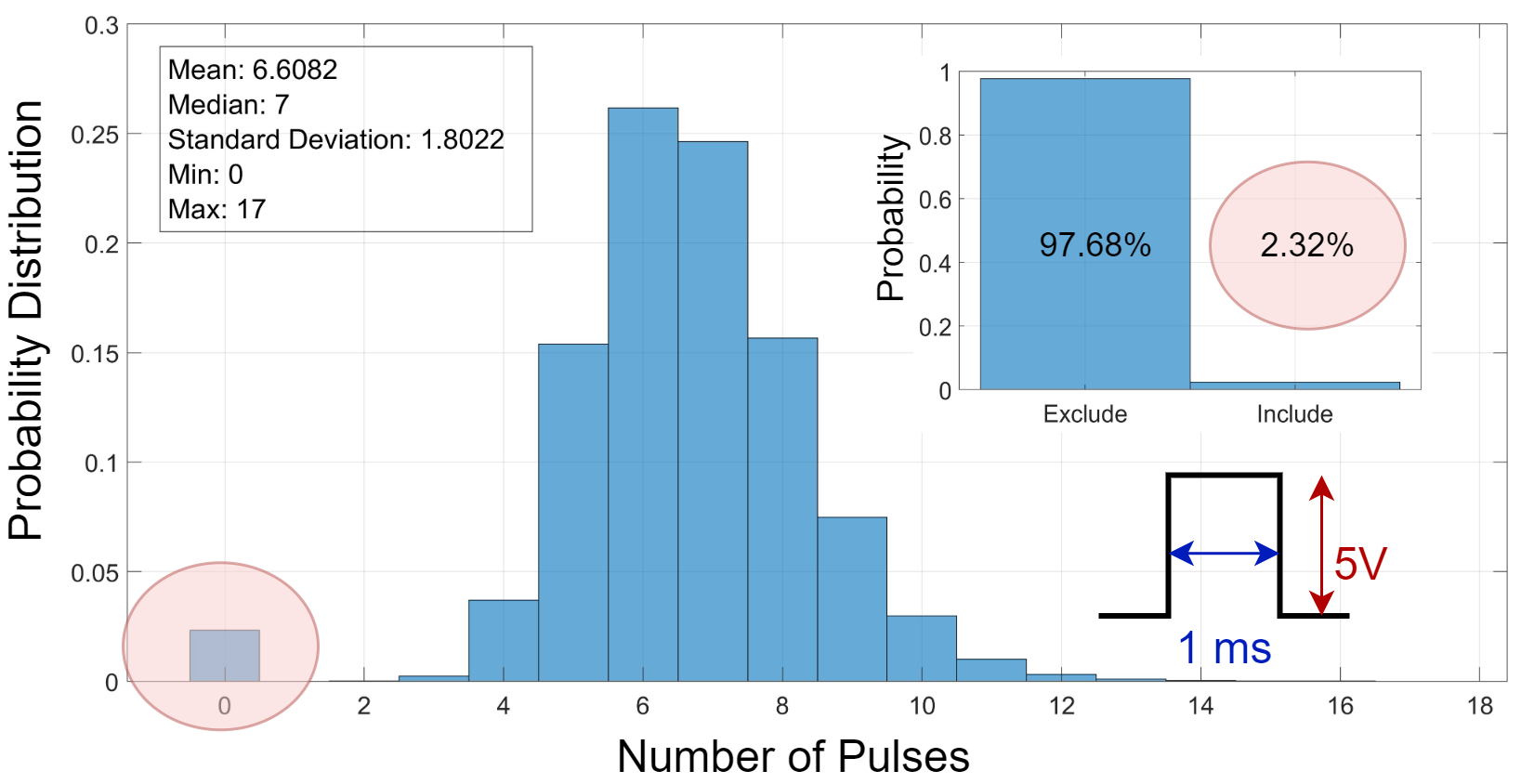}
\caption{The allocation of programming pulses for storing the actions of TA into the clause crossbar array. A pulse width of \((1\ ms)\) is used to store the exclude actions and program the cells from initial HCS \(\approx2.5\ \mu S\) to LCS \(\approx1\ nS\). The inset figure shows the percentage of the two actions over the entire array.}
\label{fig_Ta_pulse-dist}
\end{figure}
\vspace*{-5pt}

\begin{figure}[t]
\centering\includegraphics[width=\linewidth]{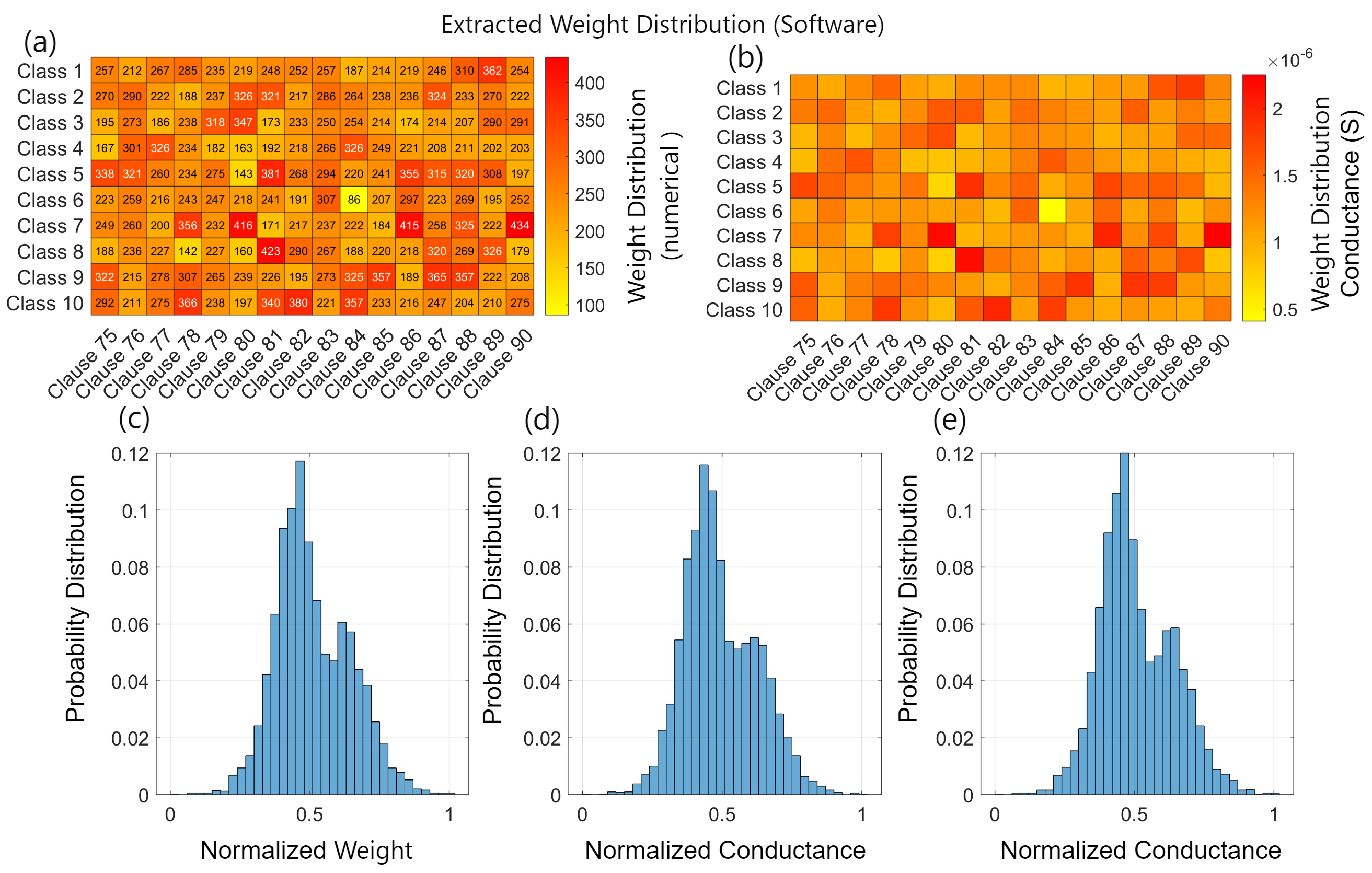}
\caption{Clause software weight matrix and class crossbar tile. (a) A sample of a unipolar software clause weight matrix for clauses ranging from 75 to 90. (b) After programming the allocated memory to store the clause weight matrix for clauses between 75 and 90 as tunable conductance values. (c) Distribution of the normalized values of all the unipolar software clause weight matrix. (d and e) Distribution of the normalized conductance values of class crossbar tile after the pre-tune and fine-tune, respectively.}
\label{fig_Weight_dist}
\end{figure}
\vspace*{-5pt}

\subsection{Encoding TAs/Weights}\label{Sec:Encoding TAs/Weights}

Figure~\ref{fig_Ta_dist} displays the encoding of the trained TA Matrix (Literals \(\times\) Clauses) of CoTM for the MNIST dataset on the clause crossbar tile. The TA states' values in the trained CoTM model range between 1 and 256 (see Figure~\ref{fig_Ta_dist}a). Once the final stable states are reached, the numerical TA values are mapped to one of two Boolean actions. An include action ``1'' is initiated if the TA value exceeds half of 128, while an exclude action ``0'' is if the TA value is less than or equal to 128 (see Figure~\ref{fig_Ta_dist}b). After that, the include and exclude actions are encoded as HCS \(\approx2.5\ \mu S\) and LCS \(\approx1\ nS\), respectively, into the clause crossbar tile. In Figure~\ref{fig_Ta_dist}c, a sample of the clause crossbar tile dedicated to clauses between 75 and 90 and their corresponding TA values from 1 to 10 before the encoding of the actions.
Furthermore, Figure~\ref{fig_Ta_dist}d displays the same array after encoding the TA's actions, where a pulse width of 1 ms is used to program the actions in order to accelerate reaching the conductance boundaries of the Y-Flash device. Figure~\ref{fig_Ta_pulse-dist} shows the main histogram illustrating the probability distribution of the number of pulses required for encoding all actions in the clause crossbar tile \((2048 \times 500)\). The mean number of pulses is approximately 7,  and an SD of 1.8. The distribution shows a minimum of 0 and a maximum of 17 pulses. The inset histogram in Figure~\ref{fig_Ta_pulse-dist} emphasizes the proportion of include and exclude actions, with \(97.68\%\) of the data points falling into the exclude category and \(2.32\%\) into the include category. This distribution shows that exclude actions are the majority in the CoTM. The low probability of needing zero pulses (highlighted in the red circle) indicates that very few TAs are set to an include state. This small ratio of included actions indicates the energy efficiency of the CoTM algorithm, as this action consumes the highest power during the inference process. By minimizing the frequency of include action, the overall power consumption is reduced, thereby enhancing the energy efficiency of the IMPACT.

For storing the weight, the signed weights of the trained CoTM module were converted (see Figure~\ref{fig_Tuning_procedure}) to unsigned (unipolar) weights ranging from 0 to 419, as illustrated in Figure~\ref{fig_Weight_dist}a. The conductance range of the Y-Flash device $(1\ nS - 2.5\ \mu S)$ was divided into uniform segments, assigning a specific target conductance value to each weight level. This conversion and segmentation ensure good granularity when encoding the weight across the class crossbar tile. Before mapping the weights to the crossbar tile, all cells in the array were erased to the HCS. This approach facilitated a uniform transition to the target segments. Figure~\ref{fig_Weight_dist}b shows a sample of the crossbar tile dedicated to weights of clauses between 75 and 90 after applying the tuning mechanism. It is clear that the class crossbar tile accurately mirrors the trained CoTM's weights but as conductance values.

\begin{figure}[t]
\centering\includegraphics[width=\linewidth]{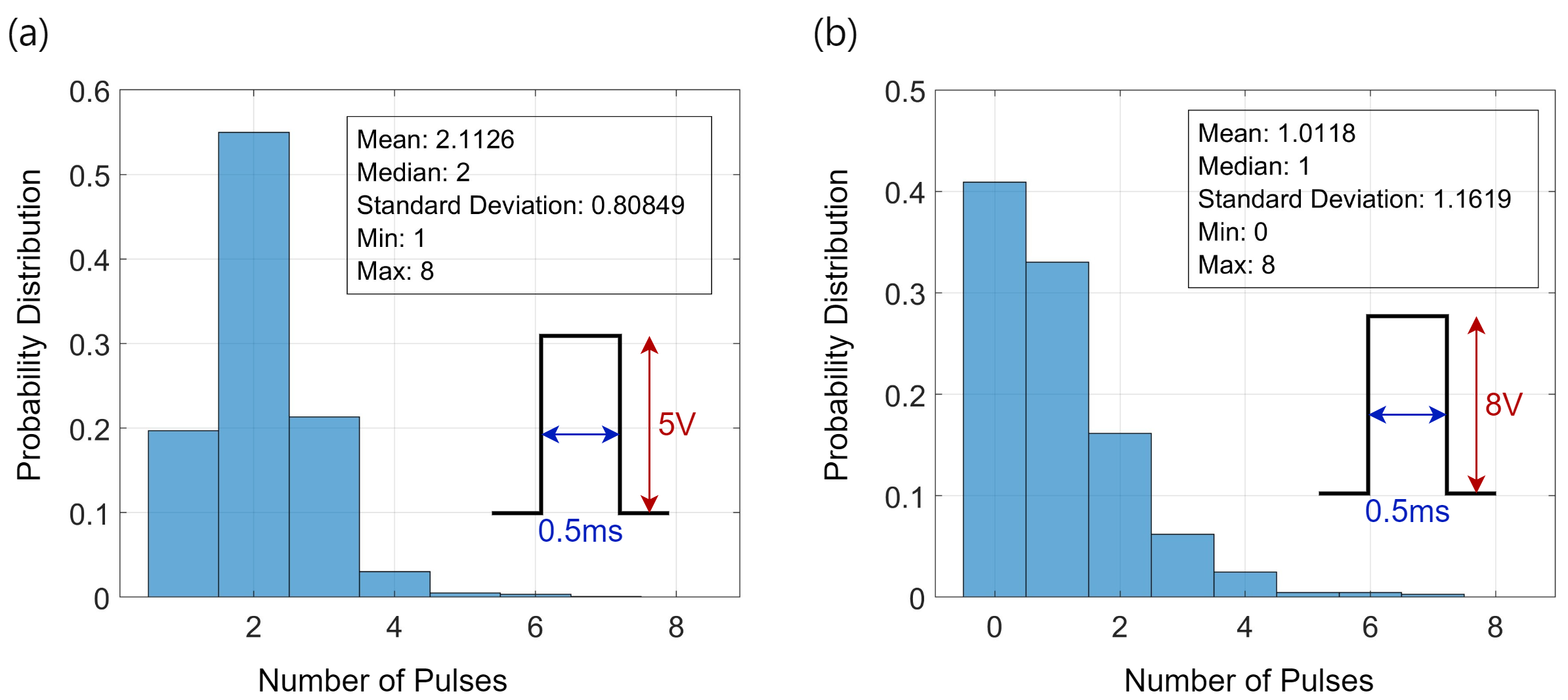}
\caption{ Store the CoTM-trained model weights in the class crossbar tile. (a) Programming pulses applied during the pre-tuning stage: \(5 \ V\) for \(0.5\ ms\) to transfer the related \(W_{ij}\)'s conductance to an LCS. (b) Erasing pulses applied during the pre-tuning stage: \(8\ V\) for \(0.5\ ms\) to erase to an HCS.}
\label{fig_w_pulse_dis}
\end{figure}

The distributions of normalized software weights, crossbar weights after pre-tuning, and crossbar weights after fine-tuning are shown in Figure~\ref{fig_Weight_dist}c, d, and e, respectively. It is evident from these distributions that there is a clear similarity between the software weights and the pre-tuned weights in the crossbar array, indicating a reliable mapping of the trained module's weights into the hardware. The conductance values of the pre-tuned crossbar weights range from a minimum of \(72.89\ nS\) to a maximum of \(2.58\ \mu S\). Similarly, after fine-tuning, the conductance values range from \(53.66\ nS\) to \(2.501\ \mu S\). As reflected in their distribution, the fine-tuned crossbar weights indicate a higher correlation with the software weights. This refinement ensures that the conductance values in the crossbar more accurately mirror the software weights, enhancing the overall performance.

The efficiency of our architecture is highlighted by the minimal pulse budget required to reach the target conductance values corresponding to the weights. Figures~\ref{fig_w_pulse_dis}a and b illustrate the distribution of the applied programming and erasing pulses in the pre-tune phase necessary to achieve these accuracy states. The range of programming pulses needed is 1 to 8 pulses, with a mean of 2 pulses and an SD of 0.8. The range for erasing pulses is 0 to 8, with a mean of 1.01 and an SD of 1.16.
This process achieved an accuracy of \(96.2\%\) for the MNIST dataset and reduced the cost error by \(0.62\%\) after a total of 10 pulses (program + erase), even without fine-tuning. During the fine-tuning phase, a maximum of 6 pulses (program + erase) was used to push the accuracy to \(96.31\%\) and reduce the \(0.62\%\) cost error to \(1\%\) with a very narrow mapping segment of \(\pm 5\).
Further, this accuracy is the same as that achieved in the software training within 25 epochs. Higher accuracy, such as \(99\%\), can be reached with the same array sizes by training for more epochs \cite{glimsdal2021coalesced}.
The minimum pulse budget required for achieving high accuracy and the resulting power efficiency underlines the distinction of IMPACT as a robust and energy-efficient architecture. This is particularly important for implementing large-scale ML, where power efficiency directly impacts the functionality and scalability of the system.

\section{Performance and Efficiency Evaluation of the IMPACT}\label{Sec:Performance and Efficiency Evaluation of the IMPACT}

For evaluating the performance and efficiency of the IMPACT architecture, we measured several key metrics: accuracy, energy consumption, area, and performance, as shown in Table \ref{Table:Power}. The MNIST dataset, comprising 10,000 test images, was used to assess the architecture, which employs Y-Flash devices at the core of its memory to implement the CoTM inferences.

The programming energy consumption/pulse was measured  \((5\ V \times 139\ \mu A\times 200\ \mu s)\). However, the current can be less than this because it depends on the state of the conductance, and this current was measured without using compliance current to limit the current where it can be limited to \(20\ \mu A\)\cite{YFLASH2022_DBNN}. The erase pulse \((8\ V\times 1\ nA \times 100\ \mu s)\)  the voltage applied to both transistors SI and SR in the two terminal configuration of Y-Flash. The reading energies in the clause crossbar tile were measured for HCS \(0.05\ pJ\) at a current of \(5\ \mu A\), and for LCS \((3.2\times 10^{-5})\ pJ\) using the average current sensed by the CSA (see Figure~\ref{fig_CSA}).

Moreover, the operation in the IMPACT architecture is defined as reading one column in the crossbar array, with each column consisting of 2048 cells. To provide a concrete example, we measured the energy consumption for reading a single column under the worst-case scenario, where all memristors are programmed to the HCS. This scenario represents the upper bound of energy demands per operation in the IMPACT system. The measurements showed that the energy consumed for reading one column is $5.76\ pJ$. This value reflects the non-linear behavior of the memristor array, where energy consumption is not simply additive. The non-linearity arises from factors such as parasitic currents, interactions between cells in the crossbar, and sneak path currents, all of which influence the overall energy per operation.

For calculating the inference energy per data point, the total consumed energy was measured, and then the average was considered for these calculations. The energy per data point was \(67.99\ pJ\) for the clause crossbar tile (array size \(500 \times 1568\)). Based on the assumption of mapping the TA in Table~\ref{Table:Ta mapping} and using the reading LCS energy to represent the exclude of literal ``0'' and the reading HCS energy to represent the include of literal ``0'', while the energy consumed when the input is literal ``1'' is negligible as the input was floating ``Z''. In the \(10 \times 500\) class crossbar tile array, the energy of each cell varies based on its weight. Accordingly, the energy was measured at a reading voltage of 2 V during the inference time for each cell, resulting in an energy consumption of \(16.22\ pJ\) per data point. Using the area \(3.159\ \mu m^2 /device\) \cite{YFLASH2022_DBNN}, the area of both computation tiles (clause and class) was calculated to be \(2.477\ mm^2\) and \(0.016\ mm^2\), respectively.

The performance in terms of GOPS (Giga Operations Per Second) was determined by evaluating the number of operations per second during clause and class computations to be 413.6. To ensure a fair comparison against different algorithms, the computation operation used here is the clause computation in the clause crossbar tile and class computation in the class crossbar tile, which is equivalent to two MAC operations in neural network algorithms.

\begin{figure}[t]
\centering\includegraphics[width=\linewidth]{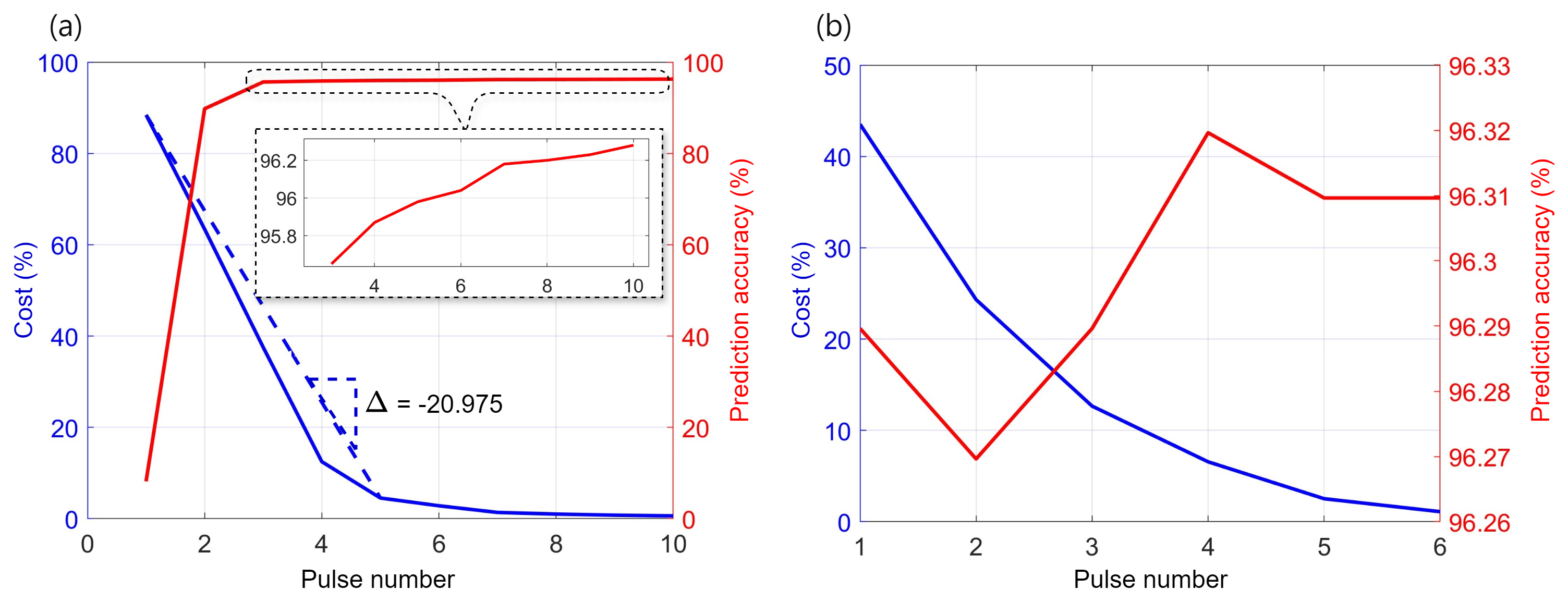}
\caption{ Evaluation of the cost and accuracy of Y-Flash crossbars using a trained CoTM on the MNIST dataset. (a) Pre-tuning stage. We observed a significant increase in accuracy with minimal pulses needed for transferring CoTM weights to their corresponding conductance in the class crossbar tile. By applying pre-tuning, notable efficiency in pulse count was achieved, with an error reduction rate of \(20.9\%\) and a cost of \(4.54\%\) after just five pulses. The cost measures the number of \(W_{ij}\) cells that failed to map the weight within the target conductance range over the total cells in the array. the pre-tuning reached an \(95.6\%\) accuracy with three pulses, increased with further pulses up to the \(10^{th}\) (see inset figure). (b) The fine-tuning stage achieved further improvements in accuracy and cost efficiency.}
\label{fig_Tunce_acc}
\end{figure}

\begin{figure}[t]
\centering\includegraphics[width=\linewidth]{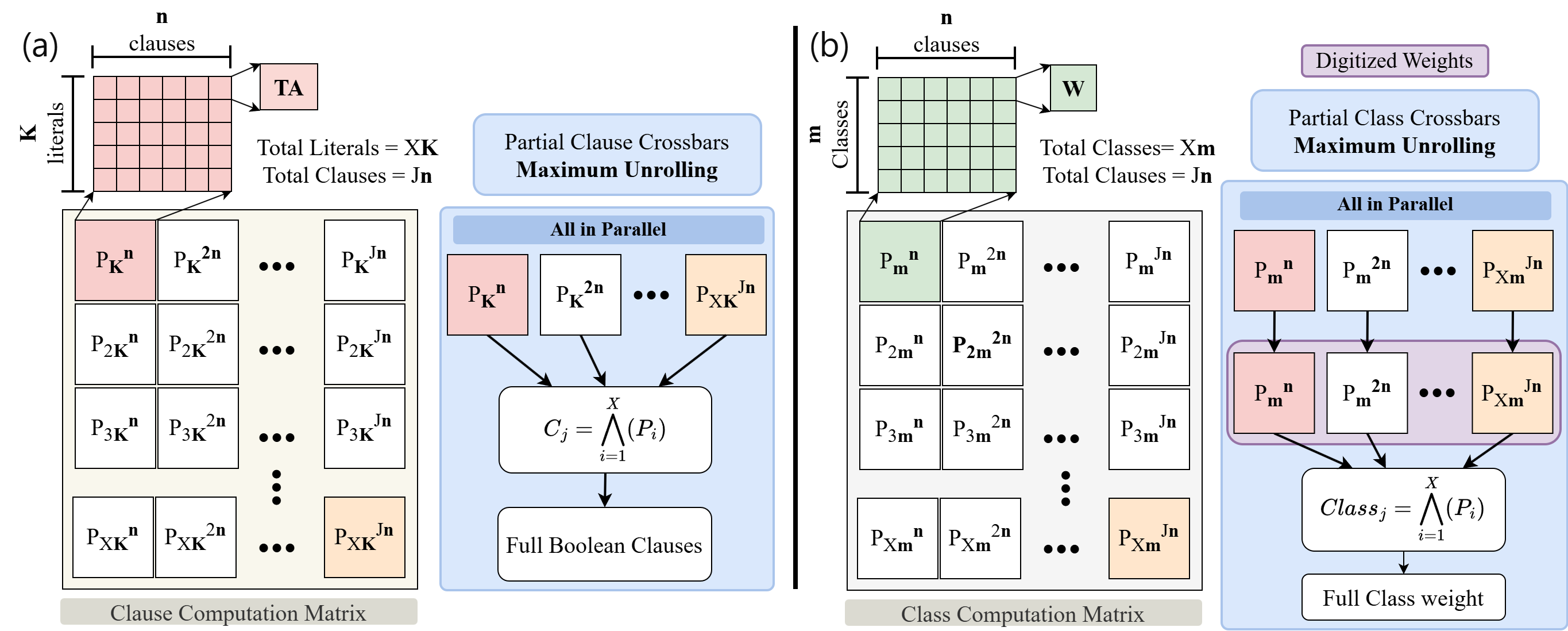}
\caption{Addressing the scalability to larger CoTM models using the modularity of the IMPACT architecture. (a) Parallel task distribution for Clause Computation Matrix. Each tile P in the matrix represents the maximum array size ($n\times K$) of the IMPACT architecture. (b) Parallel task distribution for the Class Computation Matrix. Similarly, each tile P in the matrix represents the maximum array size (n$\times$ m). All P tiles can be maximally unrolled to compute the full matrix in parallel or in subsets through a combination of temporal and spatial unrolling.}
\label{fig_Unrolling}
\end{figure}

The IMPACT architecture ensures scalability by addressing two key factors: array size limitations and task distribution across multiple arrays. The first aspect involves mitigating array size limitations by addressing the inherent electrical characteristics of memristor devices, including nonlinear voltage drops and sneak-path currents across large crossbar arrays. As the array size increases and because of these limitations, there is an increasing potential for performance degradation. However, our design incorporates techniques to mitigate these effects, such as self-selecting behavior in the Y-Flash memristors, which minimizes sneak-path currents and ensures stable performance for reasonable-sized arrays (e.g., 2048 $\times$ 500, as employed in our MNIST experiments), see section {\ref{Sec:Methods}}{\ref{section:Y-Flash Devices}}. Further, the IMPACT design specification mitigates the memristor's inherent limitations with the help of CoTM's propositional logic. It is achieved by representing a literal '1' as a floating node and thus reduces the number of active nodes in a single column during the clause computation cycle. This approach improves energy efficiency and reduces the column resistance voltage drops, contributing to the stable performance of larger arrays.

The second aspect is the ability to distribute tasks across multiple arrays. The IMPACT architecture is organized in a hierarchical crossbar structure to achieve task distribution, see section {\ref{Sec:CoTM Inference Architecture: The IMPACT}}{\ref{Class Crossbar Tile}}, and {\ref{Sec:CoTM Inference Architecture: The IMPACT}}{\ref{Clause Crossbar Tile}}. The first array serves intermediate inferences by performing the clause computations. The clauses are processed further at a higher-level crossbar to compute the classes for the final classification decision-making. Moreover, the Boolean nature of clause computation in the CoTM enhances the modularity of the IMPACT architecture, especially when dealing with larger datasets. In such cases, the number of literals exceeds the capacity of a single clause computation. However, the clause computation task can be distributed across multiple crossbars, where each crossbar processes a subset of the literals, generating partial Boolean clauses \mbox{\cite{omarImbue,10181560}}, shown in Figure{~\ref{fig_Unrolling}}a. These partial outputs are combined outside the arrays using digital AND gates to compute the related full Boolean clause. Similarly, the class computation layer can benefit from this modular approach as shown in Figure{~\ref{fig_Unrolling}}b. When the number of weights per class exceeds the capacity of a single array, the class computation can be split across multiple crossbars. Each crossbar stores and processes a subset of the weights, and the analog outputs are digitized using Analog-to-Digital Converters (ADCs). These digitized outputs are combined in the digital domain to form the final related class weight. Figure~{\ref{fig_Unrolling}} uses the maximum unrolling parallelism for the two matrices, however, this is subject to resource availability. The P tiles can be computed in a combination of temporal and spatial subsets.

\begin{table}[t]
\caption{Energy consumption and area metrics.}
\label{Table:Power}
\begin{tabular}{lll}
\hline
\textbf{Operation} & \textbf{Metrics} & \textbf{Working Mode}\\ \hline
Programming (Avg)  & 139 nJ & -\\ 
Erasing (Avg) & 0.8 pJ & - \\ 
Reading LCS  &\(3.2\ \times\ 10^{-5} \ pJ\)& Boolean mode \\ 
Reading HCS  & \(0.05\ pJ\) & Boolean mode \\ 
Energy/Datapoint & \(67.99\ pJ\) & Clause crossbar tile \(500 \times\ 1568\) \\ 
Energy/Datapoint & \(16.22\ pJ\) & Class crossbar tile \(10 \times\ 500\) \\ 
Area & \(2.477\ mm^2\) & Clause crossbar tile \(500 \times\ 1568\) \\
Area & \(0.016\ mm^2\) & Class crossbar tile \(10 \times\ 500\) \\ 
GOPS & \(413.6\) & Operation is clause (class) computation \\ 
Energy/Operation & \(5.76\ pJ\) &\makecell[l]{ Measured with all cells in the column programmed to HCS,\\ representing the upper bound of energy consumption and\\ data independent.} \\ \hline
\end{tabular}
\vspace*{-4pt}
\end{table}

\begin{table}[b]
\caption{Implementation of different datasets using the IMPACT architecture, ordered by complexity.}
\label{Table:Diff_Dataset}
\begin{tabular}{llllllll}
\hline
\textbf{Metrics\ Dataset} & \textbf{Iris} & \textbf{CIFAR2} & \textbf{KWS6} & \textbf{F. MNIST} & \textbf{EMG Data} & \textbf{Gesture Phase} & \textbf{Human Activity} \\ \hline
Classes (\#)              & 3             & 2               & 6             & 10                & 7                 & 5                             & 6                             \\
Clauses (\#)              & 12            & 1000            & 300           & 500               & 300              & 300                          & 800                          \\
Literals (\#)             & 32            & 2048            & 754           & 1568              & 192               & 424                           & 1632                          \\
Accuracy (\%)             & 96.67         & 81              & 80.3          & 84.16             & 87                & 89                            & 84                            \\ \hline
\end{tabular}
\vspace*{-4pt}
\raggedright
\textbf{Note:} (\#) The numbers of Classes, Clauses, and Literals.

\end{table}

Table {\ref{Table:Diff_Dataset}} shows the implementation metrics for different datasets using the IMPACT architecture, where the required storage capacity exceeded a single crossbar's limits for some mappings. CIFAR-2, a subset of the CIFAR-10 dataset, consists of 2 classes of images, each containing 6,000 color images of size 32 $\times$ 32 pixels. The Keyword Spotting (KWS) dataset with 6 classes (KWS-6) contains audio recordings of spoken keywords (yes, no, up, down, left, right). The Iris flower dataset includes measurements of 150 iris flowers from three species. Fashion-MNIST consists of 10 classes of grayscale images, each of size 28 $\times$ 28 pixels. Human Activity, Gesture Phase, and EMG Data for Gestures are used for more complex classification and segmentation tasks involving body movement recognition and gesture pattern analysis. By leveraging the deployment of the full clause and class computation into multiple crossbars structure, the IMPACT architecture can scale to accommodate more complex tasks and larger datasets while maintaining performance and improving computational throughput through a parallelism computation process across multiple arrays.

While MNIST was utilized as a proof-of-concept, future work will explore larger and more complex datasets, such as CIFAR-10, to validate its scalability further. Recent advancements in the Tsetlin Machine community, including the TM composites framework, have demonstrated state-of-the-art accuracy on CIFAR-10, achieving 82.8\% accuracy with enhanced hyperparameter tuning and advanced image processing techniques~\cite{gronningsaeter2024optimized}. These developments underscore the potential for IMPACT to handle more complex datasets.

Table \ref{table:comparison} provides a comparative analysis of the IMPACT architecture with other similar works. The comparison covers different technology and algorithms' performance metrics. The IMPACT architecture achieves \(96.3\%\)accuracy on the MNIST dataset, which is competitive with other state-of-the-art technologies. Utilizing the energy performance metrics of $ TOPS/W$ (Tera Operations Per Second per Watt) and $ TOPS/mm^2 $ (Tera Operations Per Second per square millimeter), the IMPACT computational efficiency and energy efficiency are emphasized. IMPACT outperforms other technologies in terms of $TOPS/W$ by a factor of \( X \).

\begin{sidewaystable}

\renewcommand{\arraystretch}{1.5} 

\caption{Comparison of various memory technologies and algorithms with the Implementation of the IMC approach.}

\label{table:comparison}
\centering
\begin{tabular}{llllllllll}
\toprule
\textbf{Reference} & \textbf{This work} & \cite{yao2020fully} & \cite{8237202} & \cite{8579538} & \cite{joshi2020accurate} &\cite{10067610}&	\cite{10067339}&\cite{10631408}	&	\cite{10185326}
 \\ \midrule
\makecell[l]{\textbf{Computing} \\ \textbf{device} }& Y-Flash & ReRAM & NOR-Flash & SRAM & PCM &ReRAM&	STT-MRAM&	STT-MRAM&	ReRAM
\\ 
\textbf{CMOS(nm)}  & 180 & - & 180 & 65 & - &22&		28	&22&	28
\\ 
\makecell[l]{\textbf{Cell} \\ \textbf{structure}} & 1R & 1T1R & R & 10T-SRAM & 1T1R&1T1R		&1T1M	&1T1M	&1T1R
 \\ 
\textbf{Weight size}& 1 Cell & 2 Cell & 1 Cell & 1 bit & 2 Cell& Multi-mode\(^a\)& 2T2M	&Multi-mode\(^a\)	&1 Cell
 \\ 
\makecell[l]{\textbf{Algorithm}} & CoTM & CNN & \makecell{Neuromorphic\\(2 layers)} & BCNN & DNN &CNN&	BNN&	CNN	&CNN
\\ 

\textbf{Dataset}& MNIST & MNIST & MNIST & MNIST & CIFAR-10&CIFAR-10	&MNIST&	CIFAR-100	&CIFAR10
 \\ 
\textbf{Accuracy(\%)}  & 96.3 & 96.1 & 94.7 & 98.3 & 93.7 &91.9	&96.2&	-	&92.1
\\ 
\makecell[l]{\textbf{Area (\(mm^{2}\))} \\ \textbf{/Capacity}} & 3.159\(^b\) & 0.25\(^b\) & 1 & 0.063/16Kb & 0.57/256Kb &24.48/4MB &4.5/2Mb&	38.93/47.25Mb	&6/4Mb
 \\ 
 
\textbf{Switch mode}& (1000/1)M\(\Omega\) & (0.5/0.05)M\(\Omega\) & (100n/10p)A & DAC/ADC\(^c\) & \(0 - 25 \mu S\)& \makecell{LRS/HRS\\MLC}
 & LRS/HRS	&LRS/HRS	&LRS/HRS
 \\ 
\textbf{Retention} & > 10 years & - & > 10 years & - & -& - &  - & -&- \\ 
\textbf{Endurance} & \(> 10^5\) & - & \(10^5\) & - & \(> 10^6\) & - &  - & -&-\\ 
\textbf{Reading (V)} & 2 & 0.2 & 4.2 & 0.8 & 0.3&0.7 -0.8&	1-1.8&	0.65-0.8	&0.7- 0.8
 \\ 
\textbf{P/E/Reset} & 5/8 (V) & 2/4.7(V) & 1.8 - 4.2 (V) & 1(V) & Imax(450 \(\mu A\))&1.2-3.5 (V) &	-&-&-
 \\ 
\textbf{TOPS/mm$^2$} & 0.17 & 1.164 & - & - & -&0.284& -&0.018&	0.056
 \\ 
\textbf{TOPS/W} & 24.56 & 11.014 & 10 & 40.3 & 11.9&51.4	&35.2	&21.4	&27.2\\
\textbf{J/Operation} & 5.76 pJ/Column\(^d\) & 0.09 pJ/MAC & 0.2 pJ/MAC & 0.024 pJ/MAC & 0.08 pJ/MAC &	19.46 pJ/MAC &	0.028 pJ/ XNOR& 0.046 pJ/MAC	& 0.036 pJ/MAC
 \\ \bottomrule
\multicolumn{9}{l}{Note: \(^a\) Single level cell or mulit level cell. \(^b\) The area is \(\times 10^{-6} \, mm^{2}/\text{cell}\).\(^c\) The resolution of ADC and DAC. \(^d\) The size of the column is 2048 cells. } \\
\end{tabular}

\end{sidewaystable}

\begin{itemize}
  \item 2.23X compared to \cite{yao2020fully} (using CNN, and ReRAM accelerator.),
  \item 2.46X compared to \cite{8237202} (using Neuromorphic (2 layers), and NOR-Flash accelerator),
  \item 0.61X compared to \cite{8579538} (using BCNN, SRAM accelerator)
  \item 2.06X compared to \cite{joshi2020accurate} (using BCNN, PCM accelerator).
\end{itemize}

While SRAM shows higher operational metrics, the SRAM in \cite{8579538} is implemented using 65 nm CMOS technology, which is more advanced compared to the 180 nm CMOS technology used in IMPACT. Advanced semiconductor nodes generally offer enhanced performance and lower power consumption owing to their smaller feature sizes and improved transistor characteristics. Additionally, SRAM demonstrates rapid switching speeds due to its volatile nature, which avoids the need for charge retention mechanisms. This makes SRAM ideal for applications where speed is critical, although it sacrifices non-volatility. The non-volatility and compact size of IMPACT's Y-Flash memory make it an attractive option for low-power, high-performance applications, demonstrating improvements in energy efficiency over other technologies, except when compared to SRAM in certain speed-critical applications. However, the non-volatility is important for enabling on-edge ML accelerators to function independently and dependably. It also contributes to decreased power usage through optimized data synchronization, facilitates scalability in resource-constrained environments, and enhances resilience to potential interruptions. These qualities are essential for preserving accuracy and mitigating system failures. Thus, the IMPACT architecture achieves a balancing performance and efficiency. 

\section{Conclusions} 

We introduced an in-memory computing inference architecture for the coalesced Tsetlin machine algorithm (CoTM), named IMPACT. It is constructed by two memory arrays: the clause crossbar tile and the class crossbar tile. The clause crossbar tile facilitates the storage and interaction of TA actions with corresponding input Boolean literals, thereby enabling the computation of Boolean clauses. The class crossbar tile stores weights and executes the weighted sum operations by integrating these weights with the computed Boolean clauses. The CoTM principle aims to learn patterns using simple propositional logical expressions to supervise decision-making. This highlights the structural simplicity of CoTM and ensures scalability and efficiency within the IMC hierarchical inference approach. The IMPACT integrates a Y-Flash memristor device manufactured using a 180 nm CMOS process. It has a fast read speed 5 ns, data retention > 10 years, and endurance of \(10^5\) cycles. Y-Flash is used in the clause crossbar tile to operate in Boolean mode by storing a \(HCS \approx 2.5\ \mu S\) to represent the include action and a \(LCS \approx1\ nS\) for the exclude action. In the class crossbar tile, Y-Flash operates in analog tunable conductance mode. The entire conductance range \((1\ nS - 2.5\ \mu S)\) is uniformly segmented to target conductances equal to the number of the highest unsigned software weight. Furthermore, C2C and D2D tests were conducted to evaluate the variability of the Y-Flash device, demonstrating its reliability for uniform weight mapping with minimal variations. Moreover, the energy per operation, \(5.76\ \text{pJ}\), was measured with all cells in the column programmed to \(HCS\), representing the upper bound of energy consumption and ensuring data independence in this calculation. We tested the IMPACT architecture on the MNIST dataset and achieved an accuracy of \(96.3\%\) through extensive validations. We demonstrated the performance of the inference accelerator at $TOPS/mm^2 = 0.17$ and $TOPS/W = 24.56$.The results proved that IMPACT meets the demands of large-scale data processing and the corresponding need for an efficient edge computing architecture that can handle real-time inference and continuous model improvement. In future work, we will extend the evaluation of the IMPACT architecture to larger and more complex datasets, such as CIFAR-10 and ImageNet, to further validate its scalability and performance.
\enlargethispage{20pt}

\ethics{Insert ethics statement here if applicable.}

\dataccess{No additional data are available for this article.}

\aucontribute{O.G. was responsible for circuit design in the cadence EDA environment and taking the measurements, writing scripts to scale up and build the IMPACT architecture, conducting experiments, performing analysis, and leading the writing of the paper. W.W. and S.K. were responsible for taking measurement readings from the Y-Flash chips at Technion—Israel Institute of Technology, Haifa, Israel. F.M. contributed resource support, scientific discussions, and suggestions, reviewed and edited the writing. A.Y. and R.S. co-supervised the circuit and system level work of O.G., contributed to writing and editing, and supervised the entire project.}

\competing{The authors declare that they have no competing interests.}

\funding{This research received no specific grant from any funding agency in the public, commercial, or not-for-profit sectors.}

\ack{The authors thank Mr. Tosuif Rahman for his assistance with the CoTM algorithm script and Mr. Tian Lan and Mr. Komal Krishnamurthy for their valuable suggestions and scientific discussions during the stages of this work, Microsystems Group, Newcastle University, UK.}

\disclaimer{Insert disclaimer text here if applicable.}

\bibliographystyle{IEEEtran}
\bibliography{RSTA_Author_tex}

\begin{thebibliography}{10}
\providecommand{\url}[1]{#1}
\csname url@samestyle\endcsname
\providecommand{\newblock}{\relax}
\providecommand{\bibinfo}[2]{#2}
\providecommand{\BIBentrySTDinterwordspacing}{\spaceskip=0pt\relax}
\providecommand{\BIBentryALTinterwordstretchfactor}{4}
\providecommand{\BIBentryALTinterwordspacing}{\spaceskip=\fontdimen2\font plus
\BIBentryALTinterwordstretchfactor\fontdimen3\font minus \fontdimen4\font\relax}
\providecommand{\BIBforeignlanguage}[2]{{%
\expandafter\ifx\csname l@#1\endcsname\relax
\typeout{** WARNING: IEEEtran.bst: No hyphenation pattern has been}%
\typeout{** loaded for the language `#1'. Using the pattern for}%
\typeout{** the default language instead.}%
\else
\language=\csname l@#1\endcsname
\fi
#2}}
\providecommand{\BIBdecl}{\relax}
\BIBdecl

\bibitem{sebastian2020memory}
A.~Sebastian, M.~Le~Gallo, R.~Khaddam-Aljameh, and E.~Eleftheriou, ``Memory devices and applications for in-memory computing,'' \emph{Nature nanotechnology}, vol.~15, no.~7, pp. 529--544, 2020.

\bibitem{krestinskaya2024neural}
O.~Krestinskaya, M.~E. Fouda, H.~Benmeziane, K.~El~Maghraoui, A.~Sebastian, W.~D. Lu, M.~Lanza, H.~Li, F.~Kurdahi, S.~A. Fahmy \emph{et~al.}, ``Neural architecture search for in-memory computing-based deep learning accelerators,'' \emph{Nature Reviews Electrical Engineering}, pp. 1--17, 2024.

\bibitem{christensen20222022}
D.~V. Christensen, R.~Dittmann, B.~Linares-Barranco, A.~Sebastian, M.~Le~Gallo, A.~Redaelli, S.~Slesazeck, T.~Mikolajick, S.~Spiga, S.~Menzel \emph{et~al.}, ``2022 roadmap on neuromorphic computing and engineering,'' \emph{Neuromorphic Computing and Engineering}, vol.~2, no.~2, p. 022501, 2022.

\bibitem{10473924}
S.~Singh, C.~K. Jha, A.~Bende, V.~Rana, S.~Patkar, R.~Drechsler, and F.~Merchant, ``{MemSPICE}: Automated simulation and energy estimation framework for {MAGIC}-based logic-in-memory,'' in \emph{2024 29th Asia and South Pacific Design Automation Conference (ASP-DAC)}, 2024, pp. 282--287.

\bibitem{ISSCC_10067643}
F.~Conti, D.~Rossi, G.~Paulin, A.~Garofalo, A.~Di~Mauro, G.~Rutishauer, G.~m. Ottavi, M.~Eggimann, H.~Okuhara, V.~Huard, O.~Montfort, L.~Jure, N.~Exibard, P.~Gouedo, M.~Louvat, E.~Botte, and L.~Benini, ``22.1 a 12.4{TOPS/W} @ 136{GOPS AI-IoT} system-on-chip with 16 {RISC-V}, 2-to-8b precision-scalable {DNN} acceleration and {30\%}-boost adaptive body biasing,'' in \emph{2023 IEEE International Solid-State Circuits Conference (ISSCC)}, 2023, pp. 21--23.

\bibitem{yu2024ferroelectric}
E.~Yu, G.~K. K, U.~Saxena, and K.~Roy, ``Ferroelectric capacitors and field-effect transistors as in-memory computing elements for machine learning workloads,'' \emph{Scientific Reports}, vol.~14, no.~1, p. 9426, 2024.

\bibitem{wan2022computeRERAM}
W.~Wan, R.~Kubendran, C.~Schaefer, S.~B. Eryilmaz, W.~Zhang, D.~Wu, S.~Deiss, P.~Raina, H.~Qian, B.~Gao \emph{et~al.}, ``A compute-in-memory chip based on resistive random-access memory,'' \emph{Nature}, vol. 608, no. 7923, pp. 504--512, 2022.

\bibitem{sebastian2019computationalPCM}
A.~Sebastian, M.~Le~Gallo, and E.~Eleftheriou, ``Computational phase-change memory: Beyond von neumann computing,'' \emph{Journal of Physics D: Applied Physics}, vol.~52, no.~44, p. 443002, 2019.

\bibitem{jung2022crossbarMRAM}
S.~Jung, H.~Lee, S.~Myung, H.~Kim, S.~K. Yoon, S.-W. Kwon, Y.~Ju, M.~Kim, W.~Yi, S.~Han \emph{et~al.}, ``A crossbar array of magnetoresistive memory devices for in-memory computing,'' \emph{Nature}, vol. 601, no. 7892, pp. 211--216, 2022.

\bibitem{Yu_9116417}
S.~Yu, A.~Soltan, R.~Shafik, T.~Bunnam, F.~Xia, D.~Balsamo, and A.~Yakovlev, ``Current-mode carry-free multiplier design using a memristor-transistor crossbar architecture,'' in \emph{2020 Design, Automation and Test in Europe Conference and Exhibition (DATE)}, 2020, pp. 638--641.

\bibitem{serb2016unsupervised}
A.~Serb, J.~Bill, A.~Khiat, R.~Berdan, R.~Legenstein, and T.~Prodromakis, ``Unsupervised learning in probabilistic neural networks with multi-state metal-oxide memristive synapses,'' \emph{Nature communications}, vol.~7, no.~1, p. 12611, 2016.

\bibitem{2018-granmo-tsetlin}
\BIBentryALTinterwordspacing
O.-C. Granmo, ``{The {T}setlin Machine - A Game Theoretic Bandit Driven Approach to Optimal Pattern Recognition with Propositional Logic},'' \emph{arXiv preprint arXiv:1804.01508}, 2018. [Online]. Available: \url{https://arxiv.org/abs/1804.01508}
\BIBentrySTDinterwordspacing

\bibitem{bao2022toward}
H.~Bao, H.~Zhou, J.~Li, H.~Pei, J.~Tian, L.~Yang, S.~Ren, S.~Tong, Y.~Li, Y.~He \emph{et~al.}, ``Toward memristive in-memory computing: principles and applications,'' \emph{Frontiers of Optoelectronics}, vol.~15, no.~1, p.~23, 2022.

\bibitem{mannocci2023memory}
P.~Mannocci, M.~Farronato, N.~Lepri, L.~Cattaneo, A.~Glukhov, Z.~Sun, and D.~Ielmini, ``In-memory computing with emerging memory devices: Status and outlook,'' \emph{APL Machine Learning}, vol.~1, no.~1, 2023.

\bibitem{lanza2021standards}
M.~Lanza, R.~Waser, D.~Ielmini, J.~J. Yang, L.~Goux, J.~Su{\~n}e, A.~J. Kenyon, A.~Mehonic, S.~Spiga, V.~Rana \emph{et~al.}, ``Standards for the characterization of endurance in resistive switching devices,'' \emph{ACS nano}, vol.~15, no.~11, pp. 17\,214--17\,231, 2021.

\bibitem{YFLASH2022_DBNN}
W.~Wang, L.~Danial, Y.~Li, E.~Herbelin, E.~Pikhay, Y.~Roizin, B.~Hoffer, Z.~Wang, and S.~Kvatinsky, ``A memristive deep belief neural network based on silicon synapses,'' \emph{Nature Electronics}, vol.~5, no.~12, pp. 870--880, 2022.

\bibitem{YFLASH2019_2terminal}
L.~Danial, E.~Pikhay, E.~Herbelin, N.~Wainstein, V.~Gupta, N.~Wald, Y.~Roizin, R.~Daniel, and S.~Kvatinsky, ``Two-terminal floating-gate transistors with a low-power memristive operation mode for analogue neuromorphic computing,'' \emph{Nature Electronics}, vol.~2, no.~12, pp. 596--605, 2019.

\bibitem{YFLASHMODEL_physical}
W.~Wang, L.~Danial, E.~Herbelin, B.~Hoffer, B.~Oved, T.~Greenberg-Toledo, E.~Pikhay, Y.~Roizin, and S.~Kvatinsky, ``Physical based compact model of {Y-Flash} memristor for neuromorphic computation,'' \emph{Applied Physics Letters}, vol. 119, no.~26, 2021.

\bibitem{glimsdal2021coalesced}
S.~Glimsdal and O.-C. Granmo, ``Coalesced multi-output tsetlin machines with clause sharing,'' \emph{arXiv preprint arXiv:2108.07594}, 2021.

\bibitem{omarImbue}
O.~Ghazal, S.~Singh, T.~Rahman, S.~Yu, Y.~Zheng, D.~Balsamo, S.~Patkar, F.~Merchant, F.~Xia, A.~Yakovlev, and R.~Shafik, ``{IMBUE}: In-memory boolean-to-current inference architecture for tsetlin machines,'' in \emph{2023 IEEE/ACM International Symposium on Low Power Electronics and Design (ISLPED)}, 2023, pp. 1--6.

\bibitem{Redress}
S.~Maheshwari, T.~Rahman, R.~Shafik, A.~Yakovlev, A.~Rafiev, L.~Jiao, and O.-C. Granmo, ``{REDRESS}: Generating compressed models for edge inference using tsetlin machines,'' \emph{IEEE Transactions on Pattern Analysis and Machine Intelligence}, vol.~45, no.~9, pp. 11\,152--11\,168, 2023.

\bibitem{10181560}
O.~Ghazal, G.~Maot, T.~Lan, J.~Ojukwu, F.~Xia, A.~Yakovlev, and R.~Shafik, ``Asynchronous control for tsetlin machine with binary memristor-transistor array,'' in \emph{2023 IEEE International Symposium on Circuits and Systems (ISCAS)}, 2023, pp. 1--5.

\bibitem{gronningsaeter2024optimized}
Y.~Gr{\o}nnings{\ae}ter, H.~S. Sm{\o}rvik, and O.-C. Granmo, ``An optimized toolbox for advanced image processing with tsetlin machine composites,'' \emph{arXiv preprint arXiv:2406.00704}, 2024.

\bibitem{yao2020fully}
P.~Yao, H.~Wu, B.~Gao, J.~Tang, Q.~Zhang, W.~Zhang, J.~J. Yang, and H.~Qian, ``Fully hardware-implemented memristor convolutional neural network,'' \emph{Nature}, vol. 577, no. 7792, pp. 641--646, 2020.

\bibitem{8237202}
F.~Merrikh-Bayat, X.~Guo, M.~Klachko, M.~Prezioso, K.~K. Likharev, and D.~B. Strukov, ``High-performance mixed-signal neurocomputing with nanoscale floating-gate memory cell arrays,'' \emph{IEEE Transactions on Neural Networks and Learning Systems}, vol.~29, no.~10, pp. 4782--4790, 2018.

\bibitem{8579538}
A.~Biswas and A.~P. Chandrakasan, ``{CONV-SRAM}: An energy-efficient {SRAM} with in-memory dot-product computation for low-power convolutional neural networks,'' \emph{IEEE Journal of Solid-State Circuits}, vol.~54, no.~1, pp. 217--230, 2019.

\bibitem{joshi2020accurate}
V.~Joshi, M.~Le~Gallo, S.~Haefeli, I.~Boybat, S.~R. Nandakumar, C.~Piveteau, M.~Dazzi, B.~Rajendran, A.~Sebastian, and E.~Eleftheriou, ``Accurate deep neural network inference using computational phase-change memory,'' \emph{Nature communications}, vol.~11, no.~1, p. 2473, 2020.

\bibitem{10067610}
W.-H. Huang, T.-H. Wen, J.-M. Hung, W.-S. Khwa, Y.-C. Lo, C.-J. Jhang, H.-H. Hsu, Y.-H. Chin, Y.-C. Chen, C.-C. Lo, R.-S. Liu, K.-T. Tang, C.-C. Hsieh, Y.-D. Chih, T.-Y. Chang, and M.-F. Chang, ``{A Nonvolatile Al-Edge Processor with 4MB SLC-MLC Hybrid-Mode ReRAM Compute-in-Memory Macro and 51.4-251TOPS/W},'' in \emph{2023 IEEE International Solid-State Circuits Conference (ISSCC)}, 2023, pp. 15--17.

\bibitem{10067339}
H.~Cai, Z.~Bian, Y.~Hou, Y.~Zhou, J.-l. Cui, Y.~Guo, X.~Tian, B.~Liu, X.~Si, Z.~Wang, J.~Yang, and W.~Shan, ``{33.4 A 28nm 2Mb STT-MRAM Computing-in-Memory Macro with a Refined Bit-Cell and 22.4 - 41.5TOPS/W for AI Inference},'' in \emph{2023 IEEE International Solid-State Circuits Conference (ISSCC)}, 2023, pp. 500--502.

\bibitem{10631408}
D.-Q. You, W.-S. Khwa, J.-J. Wu, C.-J. Jhang, G.-Y. Lin, P.-J. Chen, T.-C. Chiu, F.-Y. Chen, A.~Lee, Y.-C. Hung, C.-C. Lo, R.-S. Liu, C.-C. Hsieh, K.-T. Tang, Y.-D. Chih, T.-Y.~J. Chang, and M.-F. Chang, ``{A 22nm Nonvolatile AI-Edge Processor with 21.4TFLOPS/W using 47.25Mb Lossless-Compressed-Computing STT-MRAM Near-Memory-Compute Macro},'' in \emph{2024 IEEE Symposium on VLSI Technology and Circuits (VLSI Technology and Circuits)}, 2024, pp. 1--2.

\bibitem{10185326}
T.-H. Wen, J.-M. Hung, H.-H. Hsu, Y.~Wu, F.-C. Chang, C.-Y. Li, C.-H. Chien, C.-I. Su, W.-S. Khwa, J.-J. Wu, C.-C. Lo, R.-S. Liu, C.-C. Hsieh, K.-T. Tang, M.-S. Ho, Y.-D. Chih, T.-Y.~J. Chang, and M.-F. Chang, ``{A 28nm Nonvolatile AI Edge Processor using 4Mb Analog-Based Near-Memory-Compute ReRAM with 27.2 TOPS/W for Tiny AI Edge Devices},'' in \emph{2023 IEEE Symposium on VLSI Technology and Circuits (VLSI Technology and Circuits)}, 2023, pp. 1--2.

\end{thebibliography}

\end{document}